%% file: paper.tex
\documentclass[conference]{IEEEtran}
\let\labelindent\relax
\usepackage{floatrow}
\usepackage{enumitem}
\usepackage{color}
\usepackage{setspace} 
\usepackage[utf8]{inputenc}
\usepackage{mathtools}
\usepackage{amsthm}
\usepackage{textcomp}
\usepackage{bm}
\usepackage{amsmath}
\usepackage[noend]{algpseudocode}
\usepackage{array}
\usepackage{amssymb}
\usepackage{booktabs}
\usepackage{xspace}
\usepackage[hyphens]{url} %
\usepackage{csquotes}%
\usepackage{graphicx}
\usepackage[font=small]{caption}
\usepackage{subcaption}
\usepackage{multirow}
\usepackage{cleveref}
\usepackage{bbm}
\usepackage{tabularx} %
\usepackage[textsize=tiny, textwidth=1.1cm]{todonotes}
\usepackage{pifont}
\usepackage{caption}
\usepackage[noadjust]{cite}
\usepackage[section]{placeins}
\usepackage[linesnumbered,ruled, vlined]{algorithm2e}
\begin{document}

\makeatletter
\newcommand{\newlineauthors}{%
  \end{@IEEEauthorhalign}\hfill\mbox{}\par
  \mbox{}\hfill\begin{@IEEEauthorhalign}
}
\makeatother

\newcommand{\pluseq}{\mathrel{+}=}

\setlength{\belowcaptionskip}{-5pt}

\newcommand{\mi}{\boldsymbol{-} \mathrel{\mkern -16mu} \boldsymbol{-}}
\newcommand{\cmark}{\text{\ding{51}}}
\newcommand{\xmark}{\text{\ding{55}}}

\newcommand{\vendor}{merchant}
\newcommand{\vendors}{merchants}
\newcommand{\vendorCap}{Merchant}
\newcommand{\vendorsCap}{Merchants}

\newcommand{\ccurr}{c\kern-2.3pt$\mathrel{\text{$c$\llap{$\varparallel$}}}$} 

\newcommand{\currency}{\crypto\xspace}
\newcommand{\name}{\textit{PayPlace}\xspace}
\newtheorem{thm}{Theorem}[]
\newtheorem{lem}{Lemma}
\newtheorem{prop}[thm]{Proposition}
\newtheorem{cor}{Corollary}
\newenvironment{sproof}{%
  \renewcommand{\proofname}{Proof Sketch}\proof}{\endproof}

\newtheorem{defn}{Definition}
\newcolumntype{M}{>{\centering\arraybackslash}m{2cm}}
\newcolumntype{L}{>{\centering\arraybackslash}m{15cm}}

\setlist{leftmargin=4mm}
\newenvironment{tightenum}{\begin{itemize}\setlength{\itemsep}{0pt}\setlength{\parskip}{-1pt}\setlength{\parsep}{-1pt}}{\end{itemize}}

\newcommand{\change}[2]{\textcolor{red}{\sout{#1}}\textcolor{blue}{#2}}

\renewcommand{\change}[2]{#2}

\def\crypto{%
  \leavevmode
  \vtop{\offinterlineskip %
    \setbox0=\hbox{c}%
    \setbox2=\hbox to\wd0{\hfil\hskip-.03em
    \vrule height .3ex width .15ex\hskip .08em
    \vrule height .3ex width .15ex\hfil}
    \vbox{\copy2\box0}\box2}}
    
\pagestyle{plain} %

\title{\name: Secure and Flexible Operator-Mediated Payments in Blockchain Marketplaces at Scale}
\author{
  \IEEEauthorblockN{Madhumitha Harishankar}
  \IEEEauthorblockA{\textit{ECE Dept., Carnegie Mellon University}\\
  mharisha@andrew.cmu.edu}
  \and
  \IEEEauthorblockN{Dimitrios-Georgios Akestoridis}
  \IEEEauthorblockA{\textit{ECE Dept., Carnegie Mellon University}\\
  akestoridis@andrew.cmu.edu}
  \and
  \IEEEauthorblockN{Sriram V. Iyer}
  \IEEEauthorblockA{\textit{Flipkart}\\
  sriramv.iyer@flipkart.com}
  \newlineauthors
  \IEEEauthorblockN{Aron Laszka}
  \IEEEauthorblockA{\textit{CS Dept., University of Houston}\\
  alaszka@central.uh.edu}
  \and
  \IEEEauthorblockN{Carlee Joe-Wong}
  \IEEEauthorblockA{\textit{ECE Dept., Carnegie Mellon University}\\
  cjoewong@andrew.cmu.edu}
  \and
  \IEEEauthorblockN{Patrick Tague}
  \IEEEauthorblockA{\textit{ECE Dept., Carnegie Mellon University}\\
  tague@cmu.edu}
}
\maketitle

\begin{abstract}
Decentralized marketplace applications demand fast, cheap and easy-to-use cryptocurrency payment mechanisms to facilitate high transaction volumes. %
The standard solution for off-chain payments, state channels, are optimized for frequent transactions between two entities and impose prohibitive liquidity and capital requirements on payment senders for marketplace transactions. %
We propose \name{}, a scalable off-chain protocol for  payments between consumers and sellers. Using \name{}, consumers establish a virtual unidirectional payment channel with an intermediary operator %
to pay for their transactions. %
Unlike state channels, however, the \name{} operator can reference the custodial funds accrued off-chain in these channels to in-turn make tamper-proof off-chain payments to \vendors{}, \textit{without locking up corresponding capital in channels with \vendors{}}. %
Our design ensures that new payments made to \vendors{} are %
guaranteed to be safe once notarized %
and provably mitigates well-known drawbacks in previous constructions like the data availability attack and ensures that neither consumers nor \vendors{} need to be \textit{online} to ensure continued safety of their notarized funds.
We show that the on-chain monetary and computational costs for \name{} is $\boldsymbol{O(1)}$ in the number of payment transactions processed, and is near-constant in other parameters in most scenarios. \name{} can hence scale the payment throughput for large-scale marketplaces at \textit{no marginal cost} and is orders of magnitude cheaper than the state-of-art solution for non-pairwise off-chain payments, Zero Knowledge Rollups.
\end{abstract}

\setlength{\marginparwidth}{1.375cm}

\input{Introduction.tex}
\input{SystemModel.tex}
\input{Architecture.tex}
\input{ProtocolSpecification.tex}
\input{Eval.tex}
\input{Discussion.tex}

\input{Conclusion.tex}

\clearpage
\bibliographystyle{IEEEtran}
\bibliography{references}
\clearpage
\input{Appendix.tex}
\end{document}

%% file: Introduction.tex
\section{Introduction}
\label{sec:intro}

Facilitating fast and cheap cryptocurrency payments is important for several marketplace applications that use blockchains. For instance, many blockchain networks~\cite{orchid, nodle, helium} aim to facilitate sharing of last-mile network resources like bandwidth and compute %
wherein consumers make frequent incremental payments to service providers for incremental resources consumed. %
There is also increasing interest in enabling well-established two-sided marketplaces like Amazon and Uber on blockchains~\cite{forbes1, forbes2}, which requires a scalable mechanism for consumers to make cryptocurrency payments to \vendors{}. %
Since blockchain transactions are known to be limited by long finality times, low throughput, and high fees~\cite{vukolic2015quest}, off-chain payment mechanisms have come to be regarded as a promising alternative. However, predominant solutions~\cite{miller2017sprites, sivaraman2018routing, dziembowski2017perun} rely on state channels that are optimized for frequent pairwise payments between two entities (unlike typical marketplace interactions) and impose prohibitively high capital and liquidity requirements on payment senders and intermediaries in the marketplace scenario (more in Section~\ref{sec:existingAndStrawman}). Yet other off-chain protocols for broader non-pairwise scenarios~\cite{mavroudis2020snappy, zkrollup} rely excessively on the root-chain for securing off-chain funds; the number of blockchain transactions they initiate (and often the associated on-chain computational load) scales linearly in the number of payment transactions (between consumers and \vendors{}), thereby incurring substantial transaction fees and being inherently limited by the throughput of the root-chain. %

To the best of our knowledge, no work has yet addressed these practical capital and liquidity challenges in making large quantities of consumer-\vendor{} cryptocurrency payments in limited-throughput, high-cost and resource-constrained blockchains. On the other hand, several proposals~\cite{an2019truthful, feng2018competitive, wang2018blockchain, nodle} have presumed the existence of such a mechanism to design sophisticated blockchain-federated marketplaces, e.g. for crowdsensing. %
In this work, we develop \textbf{\name, a protocol enabling flexible cryptocurrency payment schemes for large-scale marketplace applications}. \name{} takes advantage of the presence of \textbf{marketplace operators} (e.g. Uber/Amazon) that can act as dedicated intermediaries for payment transactions. Hence, \name{} does not impose excessive capital requirements on consumers; they simply pay the operator for their marketplace orders rather than establish a state channel with dedicated capital with each corresponding \vendor{}. Unlike typical payment intermediary-based routing methods, however, \name{} does \textbf{not impose any liquidity requirements on the operator}. Instead, the \name{} operator temporarily holds consumers' off-chain payments custodial and periodically makes off-chain payments to corresponding \vendors{} by directly referencing these accrued off-chain funds. For every such holding period, the operator generates a \textit{short commitment} or hash of the aggregate payments to \vendors{} and \textit{notarizes} it on the root-chain. %

This operator-mediated temporarily custodial model enables flexible payment schemes, e.g. by allowing marketplace operators to match buyers with sellers asynchronously. For instance, Amazon may decide which of multiple \vendors{} should fulfil an order well after the consumer has paid for it. The custodial holding and periodic forwarding also allows for a natural reduction in the amortized cost per payment transaction; the operator aggregates off-chain payments for multiple orders received from multiple consumers in that duration and makes just one root-chain transaction to represent the off-chain payments made in-turn to each \vendor{}.

Assuring \textit{safety} of users' funds is challenging in such protocols that involve periodic notarization by an operator of off-chain payment activity on the root-chain~\cite{gudgeon2019sok,poon2017plasma,mvp,morevp,khalil2018nocust, khalil2019system} (called commit-chains or sidechains). To minimize computational and storage resource expenditure on the root-chain, only a short commitment of the off-chain payment activity (typically an irreversible hash) between users is revealed to the smart-contract. Hence the contract often does not have the ability to assess the validity of the represented transactions and resulting balances. This threatens \textit{safety} of users' funds and is a major source of concern in \name{}. %
Indeed, a \vendor{} or operator must not be able to withdraw a larger portion of a  consumer's funds than what the consumer has already sent as off-chain payments for marketplace orders to the operator. %
Similarly, funds once assigned by the operator as payments to \vendors{} must be safeguarded from future tampering as well,
including \textit{double-spend attacks} that the operator may launch, wherein the amount assigned by the operator \vendors{} exceeds what the operator has available as off-chain payments from consumers. %

\vendorsCap{} must also be safe from \textbf{data availability} attacks~\cite{dataavailability}. %
With previously proposed commit-chains~\cite{gudgeon2019sok,poon2017plasma,mvp,morevp,khalil2018nocust, khalil2019system}, the operator could submit a commitment to the root-chain without revealing included transactions (used to generate the hash) to users. Users then cannot verify whether their transactions were included, leaving them unsure of whether the operator has included malicious transactions, whether previously assigned funds are safe, and how much they are eligible to withdraw as of the latest commitment. Neither can the smart-contract verify the validity of off-chain transactions from the (irreversible) hash it receives. This in turn \textbf{necessitates that users be online} and monitor the root-chain; %
if malicious activity like the data availability attack is detected, users are expected to initiate withdrawal of their funds, leading to the well-known problem of \textbf{mass exits}~\cite{adler2019building, dziembowski2020lower}. Expecting consumers and \vendors{} to be online, however, significantly limits the practicality of the solution, especially in retail/marketplace settings. %
\begin{figure}
    \centering
    \includegraphics[height=.5\textwidth]{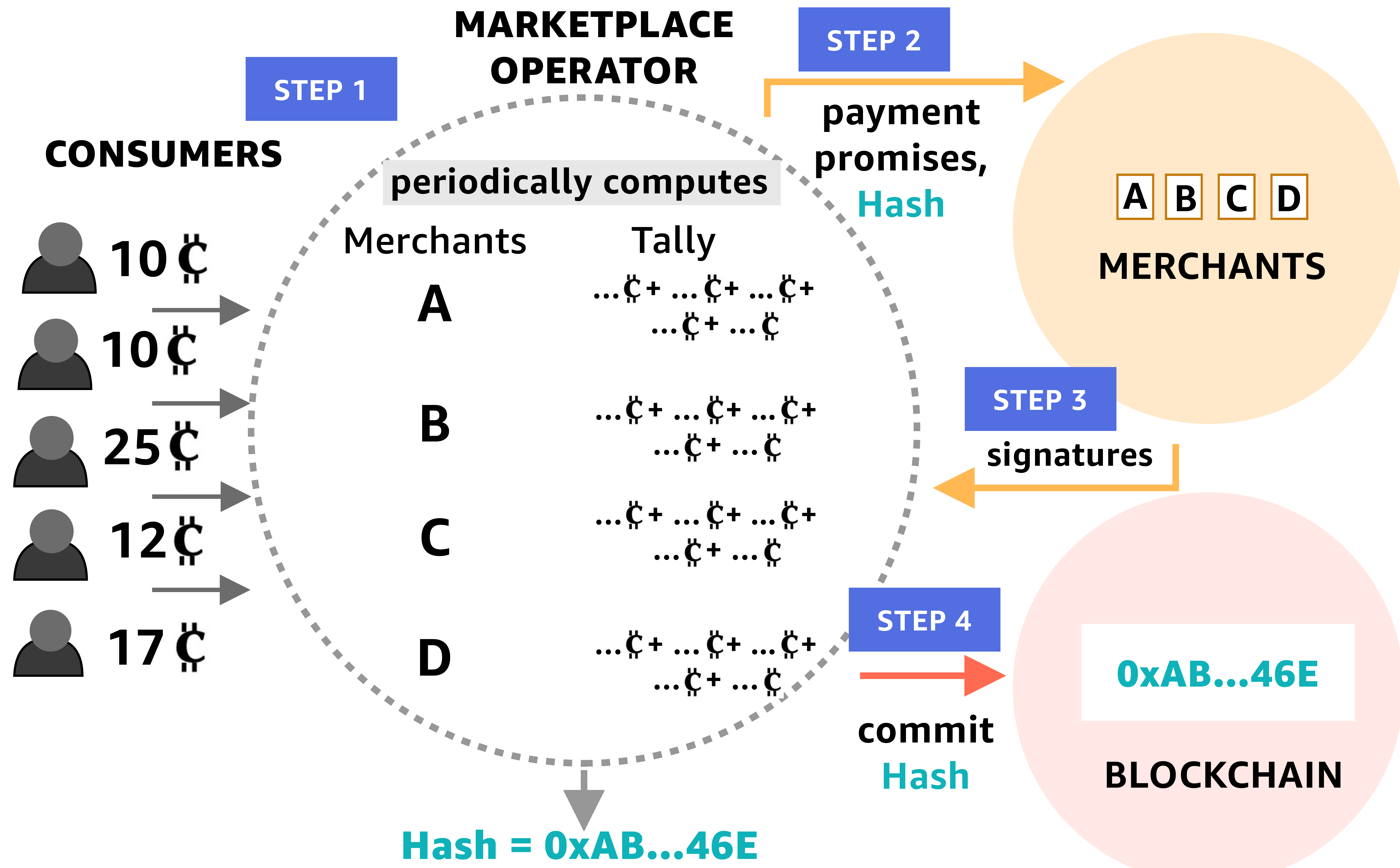}
    \caption{The \name{} operator periodically tallies the accrued consumer payments that it owes to each \vendor{}, acquires their signatures on a representative commitment, and submits it to the root-chain.}
    \label{fig:PlasmaGoSketch}
\end{figure}

\name{} solves these challenges with novel constructions tailored to the marketplace context. Figure~\ref{fig:PlasmaGoSketch} llustrates key aspects of the protocol (\currency denotes the cryptocurrency unit, e.g., ETH or BTC). %
First, we provide an easy-to-use view of the system to consumers, wherein they deposit funds in the \name{} smart-contract and regard this as a \textbf{virtual unidirectional payment channel}~\cite{spilman} with the operator. %
Consumers then make fast payments to the operator off-chain for orders placed in the marketplace without needing to be online to guard their funds. %
Second, the operator \textbf{periodically computes} payments to \vendors{} based on accrued off-chain payments, generates a short commitment or hash of this, broadcasts these computed payments and the commitment to \vendors{}, and also \textbf{reveals the off-chain funds accumulated in its virtual channels with consumers to \vendors{}}. Third, online \vendors{} attest their signatures to this commitment if they successfully verify that \textbf{no double-spend attacks have been launched} in the operator-generated payments. %
Fourth, the operator \textbf{consolidates signatures} received from \vendors{} on the generated commitment and submits it to the \name{} smart-contract for notarization. We utilize Boneh–Lynn–Shacham (BLS) signature aggregation~\cite{boneh2003aggregate} %
to \textbf{securely and efficiently combine \vendors{}' signatures of a commitment into a single signature}, avoiding the resource costs of large-scale signature verification. 
The contract hence stores only an aggregated public key of \vendors{} and notarizes a commitment if the provided aggregate signature is verifiable against the stored aggregate key. Our construction uniquely enables the contract to accept commitments even when some \vendors{}' signatures are missing \textit{and} also protect funds assigned to them in previous notarizations  \textbf{despite not storing \vendors{}' individual public keys or balance}. %
The contract ensures that the total amount withdrawn by a \vendor{} or operator against a consumer's deposit \textbf{does not exceed the funds assigned by the consumer to the operator as off-chain payments.} %
\name{} is hence resilient to data availability attacks, provides strong \vendor{} safety and \textbf{never results in mass exits} since notarized merchant funds are guaranteed to be safe even if the operator later deviates from the protocol.

We first formalize our goals and threat model  (Section~\ref{sec:model}). We then discuss related work (Section~\ref{sec:existingAndStrawman}), introduce the \name{} architecture (Section~\ref{subsec:coreConcepts}), and develop the protocol(Section~\ref{sec:main}). %
Our evaluation shows that on-chain computational and monetary costs of \name{} are orders of magnitude lower than the recently deployed state-of-the-art technique for non-pairwise off-chain payment scaling, Zero Knowledge Rollups (Section~\ref{sec:evaluation}). We finally present a discussion of limitations and concluding thoughts (Sections~\ref{sec:discussion},~\ref{sec:conclusion}).

%% file: SystemModel.tex
\section{Overview}
\label{sec:model}

\subsection{Goals}
\label{subsec:secprop}
We refer to a generic consumer by $c$ and a \vendor{} (also referred to as service provider or simply provider) by $p$.
We define \emph{confirmed funds}\change{.}{}  $f_{c,t}$ and $f_{p,t}$ as the funds available to a consumer $c$ for spending in the marketplace as of time $t$\change{.}{} and the funds available for a \vendor{} $p$
for withdrawal as of time $t$, respectively. $f_{c,t'}$ equals the deposited amount when a consumer $c$ first joins the system at time $t'$ by depositing funds into the smart-contract, and $f_{p,0}=0$ for all \vendors{} when the system is starting out at $t=0$ (i.e. no payments yet). Note that \name{} allows consumers and \vendors{} to join and leave at largely any time.
We say that an honest user (consumer or merchant) is \emph{active} at $t$, denoted as $\mathcal{A}(u,t)=1$, if she is ``online'' listening to smart-contract events and incoming messages  at $t$ and follows the protocol in response. %
Our first property ensures predictable execution time for withdrawals initiated by payment recipients:

\begin{defn}[Liveness]
A \vendor{} or an intermediary (involved in relaying consumer to \vendor{} payments) can initiate a withdrawal of their funds at any time $t$ or wait at most a predefined duration to do so. Once initiated, a withdrawal must impose no wait-times and execute to completion immediately, subject to transaction processing latency of the root chain.
\end{defn}
Liveness is not satisfied by existing commit-chain designs~\cite{morevp, mvp} %
that rely on \textit{exit games} where the smart-contract forces users to wait for a significant period of time after they initiate withdrawals in order to prevent potential attacks.
Next, it is important to ensure that %
neither consumers nor \vendors{} are at risk anytime of having funds already assigned to them %
stolen, \textit{even if they are arbitrarily inactive}.
\begin{defn} [Consumer Safety]%
For any $t' > t$, $f_{c,t'} = f_{c,t} - \sum_i \alpha_i$ where $\{\alpha_i\}$ are the values of all payments and withdrawals that consumer $c$ makes in time interval $(t, t']$.
\end{defn}

\begin{defn} [\vendorCap{} Safety]
For any $t' > t$, $f_{p,t'} \ge f_{p,t} - \sum_i \alpha_i$ where $\{\alpha_i\}$ are the values of all withdrawals that \vendor{} $p$ makes in time interval $(t, t']$.
\end{defn}
Note that \vendors{} may have accrued additional confirmed funds during $(t,t']$ from consumer payments. Indeed, our next property assures that a protocol-compliant \vendor{} %
is not affected by malicious/colluding \vendors{}. %

\begin{defn} [Income Certainty]
$\exists \theta > 0, \delta\ge0$ such that any valid payment %
initiated at time $t$ by a consumer to a \vendor{} $p$ is available as a part of $p$'s confirmed funds by $t+\theta$ if the \vendor{} and any involved intermediary $\omega$ are continuously active during $[t', t' + \delta]$  for some $t'\ge t$ (i.e. $\mathcal{A}(p, t'') = 1$ and  $\mathcal{A}(\omega, t'') = 1$ for all $t'' \in [t', t' + \delta]$).
\end{defn}

The next property provides resilience to data availability attacks that are common in commit-chains and sidechains, %
wherein users are left unsure of their available funds.

\begin{defn} [Data Availability]
\vendorsCap{} and consumers \textit{know} their confirmed funds $f_{p,t}$ and $f_{c,t}$ at any time $t$ and the information necessary to use them. 
That is, if $f_{p,t} > 0$, $\exists t' < t$ with $\mathcal{A}(p,t') = 1$ such that $p$ was notified at $t'$ of the value of $f_{p,t}$  and received necessary information to withdraw it.
If $f_{c,t} < D_{c,t}$,  $\exists t' < t$ with $\mathcal{A}(c,t') = 1$ such that $c$ is notified or aware at $t'$ of the value of $f_{c,t}$ and information to spend it.
\end{defn}

\noindent We define liquidity and root-chain footprint requirements.

\begin{defn} [Pooled Liquidity]
A consumer can initiate a valid payment at $t$ of value up to $f_{c,t}$ to any \vendor{}.
\end{defn}
\begin{defn} [Single-Source Liquidity]
An intermediary involved in relaying consumer payments to recipient \vendors{} does not need to deposit funds in the system. %
\end{defn}
In other words, consumers need not partition their funds ahead of time for use with individual \vendors{} and their capital is directly used for finishing initiated payments. %
We next define additional notation for characterizing transaction efficiency on the root-chain. Let $n_{t,t'}$ and $p_{t,t'}$ be the number of initiated consumer payments and the corresponding number of unique \vendor{} recipients during some time interval $[t, t']$, respectively. Let $r_{t,t'}$ be the number of root-chain transactions required during $[t, t']$ to complete the initiated payments (i.e. to confirm the payments available for withdrawal/reuse by recipients). 
\begin{defn} [On-Chain Efficiency]
$\exists \beta > 0,~ \delta \in [0, \beta]$ such that $\forall k=0, 1, 2, \ldots$, the protocol satisfies $r_{k\beta, (k+1)\beta} = o(n_{k\beta, (k+1)\beta})$ and $r_{k\beta, (k+1)\beta} = o(p_{k\beta, (k+1)\beta})$
as long as \vendors{} are active for $[k\beta + t, k\beta + t+\delta]$ for some $t \in [0, \beta - \delta]$ %
\end{defn}

\subsection{Threat Model and Assumptions}
\textbf{Attacker} %
\name{} aims to satisfy the goals identified in Section~\ref{subsec:secprop}. Of these, the security properties are \textit{Consumer Safety}, \textit{\vendorCap{} Safety}, \textit{Data Availability}, and \textit{Income Certainty}. Correspondingly, the key attack vectors are:%
\begin{tightenum}
    \item %
    A malicious operator may attempt to double-spend consumer payments to multiple \vendors{} or re-assign funds assigned to \vendors{} in previously notarized commitments. The operator may also attempt to withdraw more funds from a consumer's deposit than what has been assigned to the operator through the consumer's off-chain payments. These attacks would violate \textit{Consumer} and \textit{\vendorCap{}} Safety. %
    The operator may collude with \vendors{} and may also attempt to impersonate other \vendors{} (e.g. the rogue public-key attack~\cite{ristenpart2007power}) to launch these attacks. The operator may also withhold information about a submitted commitment and hence violate \textit{Data Availability}.%
    \item %
    \vendorsCap{} may collude to withdraw more funds from the \name{} contract than what has been assigned to them, thereby violating both \textit{Consumer} and \textit{\vendorCap{} Safety}.
    \item %
    \vendorsCap{} may attempt to avoid computational burden (like attesting signatures) when possible, thereby potentially violating \textit{Income Certainty} in \name{}. %
    \item %
    Malicious consumers may attempt to make invalid off-chain payments to operators or to withdraw funds already assigned to the operator. These attacks violate \textit{\vendorCap{} Safety} and often requires violating \textit{Liveness} to guard against. %
    \item %
    Even if some \vendors{} are temporarily inactive (e.g. their communication links with the operator are attacked), their already assigned funds must not be subject to risk, i.e.\textit{ \vendorCap{} Safety}, and other active \vendors{} must still be able to receive additional income, i.e. \textit{Income Certainty}.
\end{tightenum}

\textbf{Assumptions}  We assume that the root chain is secure; in other words the adversary cannot compromise execution of the \name{} smart-contract on the root-chain or impact the consensus process of root-chain miners. We also assume that the root-chain supports BLS signature verification~\cite{boneh2001short, boneh2003aggregate, boneh2018compact}. %
The BLS signature scheme and associated operations like hashing to the elliptic curve are currently being standardized~\cite{blsDraft, hashToCurve} and popular systems like Ethereum 2.0, Zcash, Chia, and Polkadot already utilize BLS signatures~\cite{blsPragmaticAggr, blsIssueZCash, chia, polkadot}. We assume that users' secret keys are secure (not leaked). Finally, we assume that the root-chain offers an inexpensive mechanism to broadcast messages and write them to logs (like Ethereum Events~\cite{ethereumevents}); the root-chain logs can be traversed to recover messages by clients who missed the broadcast. 

\section{Related Work and Strawman Designs}
\label{sec:existingAndStrawman}
\begin{table*}[]
    \centering
    \singlespacing
        \begin{tabular}{|>{\centering\arraybackslash}m{2.9cm}||>{\centering\arraybackslash}m{1.1cm}|>{\centering\arraybackslash}m{1.3cm}|>{\centering\arraybackslash}m{1.2cm}|>{\centering\arraybackslash}m{1.4cm}|>{\centering\arraybackslash}m{1.7cm}|>{\centering\arraybackslash}m{1.35cm}|>{\centering\arraybackslash}m{2.0cm}|>{\centering\arraybackslash}m{1.45cm}|}

        \hline
        Protocol & Liveness & Consumer Safety &  \vendorCap{} Safety & Income Certainty & Data Availability & Pooled Liquidity & Single-Source Liquidity & On-Chain Efficiency \\
        \hline \hline
        Blockchain Tx. & \cmark & \cmark & \cmark & \cmark &\cmark & \cmark & \cmark  & $\mi$\\ \hline
        Direct Channels & \cmark & \cmark & $\mi$ & \cmark &\cmark &  $\mi$ & \cmark  & \cmark\\
        \hline
        PCN & $\mi$ & \cmark & $\mi$ & \cmark &\cmark &  \cmark & $\mi$ & \cmark\\ \hline
        Payment Hubs & $\mi$ & \cmark & $\mi$ & \cmark &\cmark &  \cmark & $\mi$ & \cmark\\
        \hline
        Custodial Hubs &\cmark  & \cmark & \cmark  & \cmark &\cmark &  \cmark & \cmark &  $\mi$\\
        \hline
        Plasma-style CC & $\mi$ &  $\mi$  & $\mi$ & \cmark &$\mi$ & \cmark & \cmark & \cmark \\
        \hline
        Plasma CC w/ Sign. & \cmark  &  $\mi$   & \cmark  &  $\mi$ & \cmark  & \cmark & \cmark & \cmark \\
        \hline
        Snappy & \cmark &  \cmark  & \cmark &\cmark &  \cmark & \cmark &  \cmark & $\mi$\\
        \hline
        ZK Rollup & \cmark & \cmark & \cmark & \cmark &\cmark & \cmark & \cmark & $\mi$ \\
        \hline
        \textbf{\name{}} & \cmark &  \cmark & \cmark & \cmark &\cmark & \cmark & \cmark& \cmark\\
        \hline
    \end{tabular}
    \caption{\change{We summarize the properties}{Properties} provided by different cryptocurrency payment mechanisms applied to the marketplace context.}
    \label{tab:secComparison}
\end{table*}
Before proceeding to explain the \name{} protocol, we provide an at-a-glance review of how existing solutions perform in terms of meeting the goals stated above. Table~\ref{tab:secComparison} summarizes this. As a baseline, we first note that directly \textbf{using the root chain to make regular crytocurrency payments} to \vendors{} would satisfy almost all identified goals but \textit{On-Chain Efficiency}. Transactions processed on the root chain consume permanent disk space in mining nodes and also incur mining fees that can become prohibitively high during congestion periods. %
Consider, for instance, the ride-sharing economy that Uber facilitates by matching drivers and riders in the two-sided marketplace. Bitcoin and Ethereum transaction fees for July 2020 %
average approximately \$1.56 and \$1.04 respectively~\cite{bitcoinFee, ethereumFee}
, which represents a 6-10\% fee for a typical 5 km Uber ride in Switzerland of average cost \$13.90 and  44-54\% fee for a typical 5 km Uber ride in India of average cost \$1.34~\cite{uberFee}
. In comparison, credit card fees per transaction is typically 1.5-3\%. Since blockchains also have limited throughput,%
\textit{On-Chain Efficiency} is highly desirable. 

In the following review of alternate cryptocurrency payment mechanisms that have been proposed, we find that none simultaneously satisfy \textit{\vendorCap{}/Consumer Safety} and \textit{On-Chain Efficiency}; the latter requires moving payment transactions off-chain which then requires users to be \textit{online} atleast periodically to ensure that their funds are not stolen.

First, we consider consumers establishing \textbf{direct unidirectional payment channels with each \vendor{}} they transact with for frequent off-chain payments~\cite{spilman}. This crucially fails to enable \textit{Pooled Liquidity} and also violates \textit{\vendorCap{} Safety}, though it provides \textit{On-Chain Efficiency}; indeed, since consumers use the root-chain for deposit transactions very infrequently and only after making several off-chain transactions that exhaust the deposited amount, essentially any values of $\beta$ (cf. Defn.) provides \textit{On-Chain Efficiency}. %
We next consider \textbf{Payment Channels Networks} (PCN) and a specific instance of PCNs called Payment Hubs. With PCNs~\cite{miller2017sprites,sivaraman2018routing,dziembowski2017perun}, payment senders rely on non-custodial intermediaries that provide indirect routes (composed of state channels) to the payment recipient. Though this allows consumers to establish state channels (with locked-in funds) with a limited number of intermediaries in order to pay \vendors{}, significant limitations exist with this. %
First, this is not guaranteed to enable \textit{Pooled Liquidity} since the number of pairwise channels that consumers split their funds in depends entirely on the network topology. %
Second, this fails to provide \textit{Single-Source Liquidity} since consumers intrinsically rely on intermediaries' liquidity. %
In-fact, it has been observed that the resulting rapid fluctuations in intermediaries' link capacities %
makes it challenging to find routes reliably between payment senders and recipients~\cite{prihodko2016flare}. Recent empirical analysis of the Lightning Network~\cite{beres2019cryptoeconomic} further confirms that 1) ``merchant" nodes~\cite{lightningMerchant} receive $~80$\% of the off-chain payments, 2) nodes are hence forced to frequently close and rebalance their channels due to steady depletion of liquidity in the consumer$\rightarrow$\vendor{} direction~\cite{engelmann2017towards} and 3) routing intermediaries have low return on investment on their locked-in funds. %
These issues are exacerbated in the context of large-scale marketplaces with frequent payments between consumers and arbitrary \vendors{}. %

\textbf{Payment Hubs} are PCNs where an intermediary is dedicating to providing a 1-hop route between consumers and \vendors{}. In comparison with PCNs, this facilitates \textit{Pooled Liquidity} by %
allowing consumers to pool their capital (intended for use in marketplace orders) in a single unidirectional channel with the intermediary, e.g. as with Plasma Debit~\cite{debit}. However, the other challenges with PCNs carry over.%
 We also consider a \textbf{custodial version of Payment Hubs}, where the intermediary operator receives consumer payments in dedicated state channels with consumers and periodically initiates root chain transactions to make corresponding payments from to each \vendor{}. However, this violates \textit{On-Chain Efficiency} since the number of on-chain transactions grows with the number of \vendors{} (receiving payments) every period.

\textbf{Plasma-style commit-chains}~\cite{mvp,morevp, khalil2019system} involve periodic notarization of arbitrary off-chain payment activity on the root-chain by a dedicated intermediary and can be used to alleviate liquidity requirements. %
The notarization information is simply a short hash of off-chain transactions and does not allow the contract to track and validate individual transactions and resulting balances (by design, to minimize computational and storage resources consumed on the root-chain). %
This results in violation to both \textit{Consumer} and \textit{\vendorCap{}} \textit{Safety} violated as the operator may simply insert invalid/malicious transactions in a block and use it to withdraw their funds; users can protect their funds only if they are online to detect such activity and withdraw their funds in response. There is no clear definition of \textit{confirmed funds} for users such that these funds are safe even if users are arbitrarily offline. The operator may fail to reveal the set of transactions associated with a published commitment to users, violating \textit{Data Availability}. We consider a strawman modification to this called \textbf{Plasma-style commit-chains with Signature}, wherein \vendors{}' signatures are required by the contract on the commitment to ensure that they have been revealed necessary information about the corresponding Plasma block. While this ensures \textit{Data Availability}, it crucially violates \textit{Income Certainty}. A few \vendors{} withholding signatures maliciously (or even accidentally inactive) lend a commitment unfit for notarization. %
\textit{\vendorCap{} Safety} holds since \vendors{} implicitly agree on the transactions included in a block and the validity of resulting balances by unanimously attesting their signature on a notarized commitment. %
However, consumers are then subject to collusion attacks by \vendors{} and the operator, where older payments from consumers that have already been withdrawn by receiving \vendors{} may be included again a block. 
Hence, consumers' signatures are also required on commitments to ensure that they can protect themselves from such attacks (and exit games avoided), which in-turn necessitates that they be \textit{active} to secure their funds; \textit{Consumer Safety} is violated.%

We also consider \textbf{Snappy}~\cite{mavroudis2020snappy}, a protocol for marketplace payments that has been recently proposed in parallel to ours. With Snappy, consumers directly send payments to \vendors{} on the root-chain, but are unrestrained by the root-chain's transaction confirmation latency%
. %
However, atleast one root-chain transaction is made for each payment; hence Snappy does not provide \textit{On-Chain Efficiency}. Finally, we consider the state-of-the-art solution for off-chain payments that extends beyond pairwise transactions, \textbf{Zero Knowledge (ZK) Rollup}~\cite{zkrollup}. ZK Rollup is advocated by the Ethereum Foundation and have been deployed by multiple companies recently. Unlike Snappy, the number of on-chain transactions required to process initiated payments is typically much smaller than the number of such payments and the number of payment recipients, though not sublinear in growth (i.e. no \textit{On-Chain Efficiency}). By using ZK proofs to periodically assert the validity of several off-chain transactions at once on the root-chain, ZK Rollups simultaneously assure \textit{Safety} and \textit{Liveness}. %
Consumers deposit their funds in the Rollup contract, and transact with \vendors{} off-chain via a non-custodial operator, which enables \textit{Pooled Liquidity} as well as \textit{Single-Source Liquidity}. Updated account balances are explicitly revealed in the root-chain, ensuring \textit{Data Availability}.

As shown in Table~\ref{tab:secComparison}, \textbf{\name{}} satisfies all identified goals.
Using periodic operator-driven notarization on the root-chain, \name{} ensures non-revocation of off-chain payments made by the operator to \vendors{} and hence provides \textit{Merchant Safety}. 
Requiring $p$'s signature on a block for successful notarization also overcomes the \textit{Data Availability} attack and ensures that \vendors{} know and can access their confirmed funds. 
The operator makes \textit{one} root-chain transaction periodically to assign funds to \vendors{}; the computational costs of this transaction is \textit{constant} in the number of underlying transactions and at-worst, sublinear in then number of payment recipients, thereby guaranteeing \textit{On-Chain Efficiency}.  %

%% file: Architecture.tex
\section{\name{} Architecture} \label{subsec:coreConcepts}

\begin{figure}[!htp]
    \centering
    \includegraphics[width=.9\columnwidth]{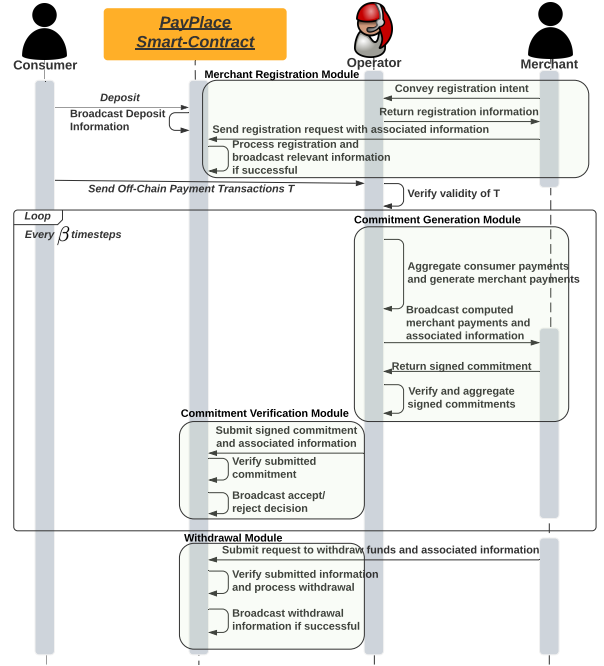}
    \caption{Sequence diagram illustrating typical interactions between Consumers, \vendorsCap{}, the \name{} contract and the Operator.}
    \label{fig:flowchart}
\end{figure}

\textbf{Consumers}
A consumer $c$ with public key $pk_{c}$ %
deposits funds in the \name{} smart-contract for making payments in the marketplace (intended for any \vendor{}). This design directly results in \textit{Pooled Liquidity}. %
The consumer may deposit more funds at any time; we use $D_{c, t}$ to denote the total funds deposited by consumer $c$ as of \change{}{time} $t$.
The \name{} smart-contract is designed to allow consumers to view their deposit as virtually establishing a \textit{unidirectional payment channel} with the operator (with refunds and returns as external to the protocol). That is, consumers make incremental off-chain payments to the operator for each order placed and need not be online to protect their unspent funds, as with unidirectional state channels. The contract's commitment verification and withdrawal modules
ensure that the total amount withdrawn by the operator and \vendors{} against a consumer's deposit does not exceed the amount assigned by the consumer to the operator, thereby facilitating this simple view.

An off-chain payment from $c$ for an order consists of a transaction $T = (\mu, pk_\omega, pk_{c})$ and the digital signature of the transaction $\sigma = \mathcal{S}(T, sk_c)$. $\mu$ is the payment amount and indicates the \emph{total} amount promised by the consumer to the operator as of when $T$ is generated, incorporating the incremental amount the consumer intends to pay for their latest order in the marketplace.
$pk_\omega$ is the operator's public key (the payment recipient), $sk_{c}$ is the consumer's private key and $\mathcal{S}(T, sk_c)$ generates a cryptographic signature using $sk_c$ on $T$.
We let $s(T)$ and $\mu(T)$ denote the sending consumer's public key $pk_c$ and the amount $\mu$ specified in transaction $T$ respectively. We use $\mu_{c,t}^\ast$ to denote the total funds spent by $c$ in off-chain payments to the operator as of $t$.
The operator verifies \change{such}{}an off-chain payment transaction $T$ received from $c$ at $t$ by evaluating if the sender has sufficient balance to make this transaction (i.e. $\mu(T) \leq D_{c,t}$) and ensuring that the operator balance in $c$'s payment channel only increases as a result of $T$ (i.e. $\mu_{c,t}^\ast \leq \mu(T)$). %
The operator also verifies the digital signature $\sigma$ with verification function $\mathcal{V}$; $\mathcal{V}(pk_c, T, \sigma)=1$ if $sk_c$ was used to sign $T$ to yield $\sigma$. We use $\mathbb{C}_t$ to denote the set of the last off-chain transaction received by the operator from each consumer as of time $t$ (reflecting the operator-owned balance in each virtual channel with a consumer as of $t$).

\begin{figure}
    \centering
    \includegraphics[width=.5\columnwidth]{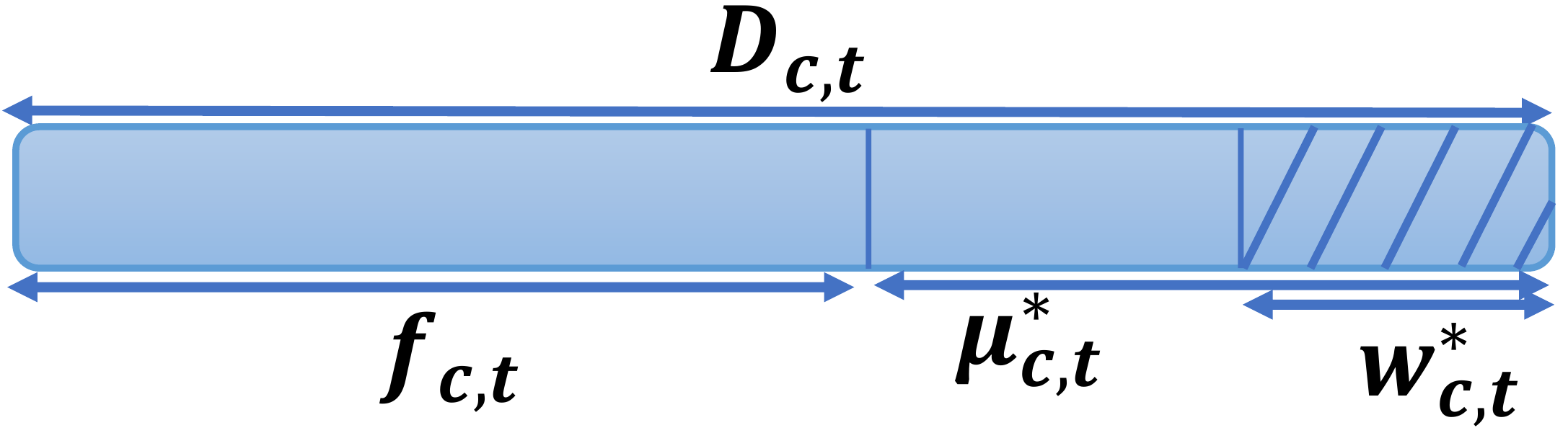}
    \caption{
    State of consumer $c$'s channel with the operator at time $t$. 
    }
    \label{fig:consumerchannel}
\end{figure}

Consumers are not permitted to withdraw funds already deposited in the channel (e.g. akin to topping up a store card). Then, a consumer's \textit{confirmed funds} at time $t$ is simply their total deposits less off-chain payments, i.e., $f_{c,t} = D_{c,t}-\mu_{c,t}^*$ (Figure~\ref{fig:consumerchannel}). We let $w_{c,t}^\ast$ denote the total funds withdrawn (by the operator and \vendors{}) against the operator-owned portion of $c$'s deposit (assigned via off-chain transactions by $c$ to the operator) as of time $t$.  We show in Section~\ref{sec:modules} that $w_{c,t}^\ast$ never exceeds $\mu_{c,t}^\ast$; i.e. \textit{Consumer Safety} is guaranteed. As shown in Figure~\ref{fig:flowchart}, for convenience, the contract may broadcast (through a mechanism like Ethereum Events) the updated value of $D_{c,t}$ (and $w_{c,t}^\ast$) when it processes a consumer deposit (or withdrawals against a consumer's funds, correspondingly). Such broadcasted data is also written out to logs. %

\textbf{Operator}
The operator holds consumers' off-chain payments custodial and forwards them \textit{off-chain} to appropriate \vendors{} every $\beta$ timesteps. Note that to start receiving payments from the operator, \vendors{} must register first by performing a one-time registration ceremony that involves the operator and the smart-contract. %
This is depicted as the \textbf{Merchant Registration Module} in Figure~\ref{fig:flowchart} which shows key entities and interactions in \name{}, and explained in detail in Section~\ref{sec:modules}.
Every $\beta$ time slots, the operator periodically consolidates payments owed to each registered \vendor{} and generates payment transactions $T'$. 
Here, $T'=(\mu', pk_p, pk_c)$ where $\mu'$ represents the \textit{total amount} owed by the operator to \vendor{} $p$ based on orders from consumer $c$ since the time $p$ last withdrew her funds on the root-chain. %
We abuse notation and use $\mu(T')$ and $s(T')$ to denote the payment amount $\mu'$ and the referenced source consumer $pk_c$ in $T'$. Every $\beta$ timesteps, a \vendor{} $p$ hence receives an off-chain transaction $T'$ for each consumer whose order(s) $p$ has fulfilled since $p$ last withdrew her funds on the root-chain; we use $\mathbb{T}_p$ to denote these transactions. 
The \name{} smart-contract allows \vendors{} to later withdraw funds assigned to them in such off-chain payment transactions $T'$ directly from the deposit of the corresponding consumer $s(T')$, thereby enabling \textit{Single-Source Liquidity}.
Note that a successful withdrawal at $t$ by a \vendor{} $p$ transfers all of $p$'s confirmed funds $f_{p,t}$ to $p$.

After computing $\mathbb{T}_p$ for all $p$, the operator generates a ``block" $\kappa=(\mathbb{T},\mathcal{M})$ that consists of the set $\mathbb{T}=\bigcup_{\forall p} \mathbb{T}_p$ %
and a Merkle tree $\mathcal{M}$. Note that $\mathbb{T}_p$ is an element of $\mathbb{T}$.
We use $\mathbb{T}(\kappa)$ to denote the set $\mathbb{T}$ in $\kappa$
and $\mathcal{M}(\kappa)$ to denote the Merkle tree $\mathcal{M}$ included in $\kappa$; $\mathcal{M}(\kappa)$ is generated from $\mathbb{T}(\kappa)$ and its root is denoted by $R(\mathcal{M(\kappa)})$.
Each leaf $L_{p}(\mathcal{M})$ in the Merkle tree $\mathcal{M}(\kappa)$ corresponds to the hash of the set of payment transactions $\mathbb{T}_p(\kappa)$ for a \vendor{} $p$ (illustration in Figure~\ref{fig:plasmablock} in Appendix~\ref{app:merkleFigure}). %
The operator also includes a similar set of transactions $\mathbb{T}_{\omega}$ assigned to herself, reflecting any commission retained from consumers' payments for providing the \name{} service. Since $\mathbb{T}_{\omega}$ is identical to any other $\mathbb{T}_p$, we do not differentiate between the operator and the \vendor{} when referring to the payees of a block, unless required. Finally, we denote the Merkle proof~\cite{halevi2011proofs} of $\mathbb{T}_p(\kappa)$ by $P_p(\mathcal{M})$, i.e. $P_p(\mathcal{M})$ proves that $\mathbb{T}_p(\kappa)$ corresponds to $L_{p}(\mathcal{M})$ and that $L_{p}(\mathcal{M})$ is a leaf of a Merkle tree with root $R(\mathcal{M(\kappa)})$. The set of \vendors{} that have leaves in $\mathcal{M}(\kappa)$ is denoted by $\mathbb{P}(\kappa)$. Appendix~\ref{sec:numericalExample} provides a numerical example illustrating off-chain payment transactions in \name{}.

\textbf{\vendorsCap{}}
The operator then broadcasts the generated block $\kappa$ to \vendors{} along with the set $\mathbb{C}_t$ %
, and the current timestamp $s_t$. Then, \vendors{} verify the block, attest their (BLS) signatures to its commitment (a hash of $R(\mathcal{M}(\kappa))$ and $s_t$) and send it to the operator. In doing so, \textit{they protect themselves from double-spend attacks by the operator}; a \vendor{} signs the root only if the operator's payments specified in $\kappa$ to \vendors{} does not exceed what the operator has been assigned from consumers. %
This directly also ensures \textit{Data Availability}; indeed, a \vendor{}'s \textit{confirmed funds} in \name{} corresponds to funds assigned to her in the last notarized commitment that she attested her signature on. The notarization process used by the \name{} smart-contract makes any incremental income specified in a notarized block inaccessible to a \vendor{} unless her signature on the corresponding Merkle root was provided by the operator during notarization. This design also incentivizes \vendors{} to be periodically \textit{active} and participate in the signing process to receive their incremental income for the last $\beta$ timesteps. Note that each commitment generated by the operator reflects cumulative payments owed to \vendors{}, hence a \vendor{} that fails to participate in one commitment round (e.g. communication links are down) can simply receive the incremental income by participating in the next round.
The operator verifies returned \vendor{} signatures on the commitment, aggregates them into a single one, %
and submits this to the smart-contract for notarization of the off-chain payments made to \vendors{} in this block. This process of computing \vendors{}' payments and acquiring their signatures is referred to as the \textbf{Commitment Generation Module} in Figure~\ref{fig:flowchart} and explained in detail in Section~\ref{sec:modules}.

\textbf{Smart-Contract} By submitting a new commitment for notarization, the operator triggers the \textbf{Commitment Verification Module} of the smart-contract. We use $\mathbb{K}_t$ to denote blocks that have been notarized as of time $t$, where $\mathbb{K}_t(-1)$ refers to the last notarized block, $\mathbb{K}_t(-2)$ to the last block and so on. If all registered \vendors{} have signed the submitted commitment, indicating that they have verified the validity of their assigned payments, the commitment is accepted. However, we must allow notarization even when some signatures are missing to ensure \textit{Income Certainty}. %
Hence, %
the \name{} contract \textit{reserves} funds that were assigned to non-signing (or ``missing") \vendors{} in the last notarized commitment $\kappa_p$ which included their signature, i.e. corresponding to $\mathbb{T}_p(\kappa_p)$.
To do this, however, it requires the operator to \textit{prove that signing \vendors} (i.e. whose signatures are included in the submitted commitment) \textit{are aware of the funds that will be set aside by the contract for non-signing ones}. %
Combined with the design of the contract's \textbf{Withdrawal Module} (explained in Section~\ref{sec:modules}), this ensures that funds assigned to a \vendor{} through a notarized commitment that contains her signature are secure, even if she subsequently becomes inactive. %
In other words, a \vendor{}'s confirmed funds $f_{p,t}$ corresponds to those assigned in $\mathbb{T}_p(\kappa_p)$. \name{} is hence impervious to the problem of \textit{mass exits}. The vendor actively participates in commitment generation only to receive \textit{additional income}. If malicious operator actions are detected (e.g. the operator fails to generate a commitment), then the \vendor{} stops fulfilling orders that are handled through this operator; a \vendor{}'s risk exposure in this semi-custodial model does not exceed the revenue of one notarization period.  %
\name{} also provides \textit{On-Chain Efficiency}. 
Only one root-chain transaction per $\beta$ is required to assign the corresponding payments to \vendors{} regardless of the number of consumer orders and recipients. %

\label{sec:timing}
\begin{figure}
    \centering
    \includegraphics[width=.9\columnwidth]{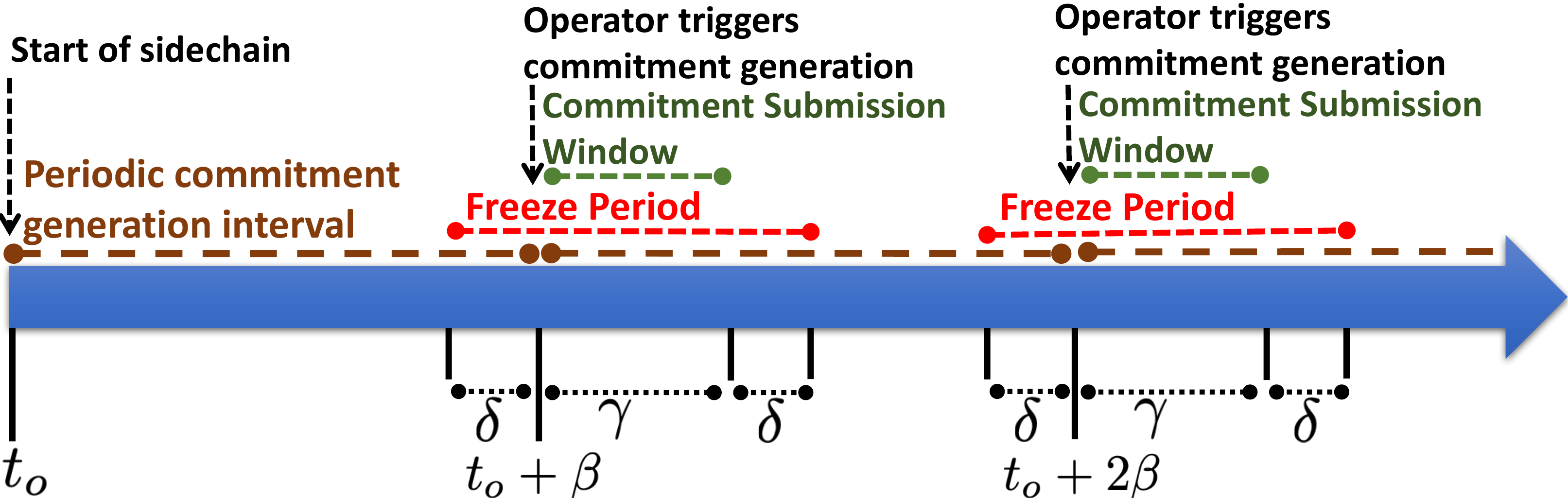}
    \caption{Depicting timing events in \name{}.}
    \label{fig:plasmaTiming}
\end{figure}
\textbf{Timing Considerations }
To protect against timing-related attacks on \textit{Consumer} and \textit{\vendorCap{} Safety}
, we introduce temporal restrictions on \vendor{} and operator actions. %
Suppose the operator starts the \name{} system at time $t_0$;  WLOG, we take $t_0=0$. %
Let $\gamma \ll \beta$ denote the (predefined) maximum time required for the operator to generate a new commitment for notarization%
, and $\delta \ll \beta$ the maximum finality time for the root blockchain. 
The smart-contract enforces \textit{freeze periods} during which funds may not be withdrawn and new \vendors{} may not register, namely, $[\beta - \delta, \beta + \gamma + \delta]$, $[2\beta - \delta, 2\beta + \gamma+\delta]$, $[3\beta - \delta,$ $3\beta + \gamma + \delta]$,~\dots The operator may submit new commitments for notarization only during the \textit{commitment submission window} within each freeze period, given by $[\beta, \beta + \gamma]$, $[2\beta, 2\beta + \gamma]$, $[3\beta, 3\beta + \gamma]$,~\dots Figure~\ref{fig:plasmaTiming} depicts the timing of these events, which can be triggered and managed using decentralized oracles~\cite{alarmclock}. A withdrawal triggered outside a freeze period is immediately processed by the \name{} smart-contract (without imposing any wait-times) as described in Section~\ref{sec:modules}, thereby providing \textit{Liveness}.

\textbf{BLS Signatures} \name{} relies on \textit{signature aggregation} to enable the contract on the root-chain to cheaply verify \vendors{}'s signatures on the submitted commitment in constant time. %
We utilize the BLS (Boneh-Lynn-Shacham) signature scheme~\cite{boneh2001short,boneh2003aggregate} based on elliptic curve pairing-based cryptography for this; it provides short signatures that can be securely aggregated. %
We let $e:\mathbb{G}_0 \times \mathbb{G}_1 \rightarrow{} \mathbb{G}_T$ denote an efficiently computable, non-degenerate pairing where $\mathbb{G}_0$, $\mathbb{G}_1$ and $\mathbb{G}_T$ are groups of prime order $q$, and $g_0$, $g_1$ are generators of $\mathbb{G}_0$, $\mathbb{G}_1$ respectively. Suppose signatures reside in $\mathbb{G}_0$ and public keys in $\mathbb{G}_1$. $H_0$ then denotes the hash function that maps from the message space into $\mathbb{G}_0$. $\mathcal{S}(m, sk)$ generates $sk$'s (BLS) signature on $m$, returning $\sigma = H_0(m)^{sk} \in \mathbb{G}_0$. $\mathcal{V}(pk, m, \sigma)$ verifies if $sk$ signed $m$ to yield $\sigma$ by evaluating if $e(g_1, \sigma) = e(pk, H_0(m))$, and returning $1$ in that case.
We use the multiplicative notation for groups, and references to PKI credentials and signatures mean BLS, unless otherwise stated. 
Using BLS signatures, only \textit{one signature verification} (two bilinear pairings) is required to check whether all required signers (represented by their aggregate public key) have signed the presented aggregate signature. \name{}'s design, however, goes further to allow the smart-contract to determine exactly which \vendors{} have not signed a commitment despite storing only an aggregated public key of registered \vendors{} (as opposed to each key individually, to save on expensive storage resources). Identifying these  non-signing \vendors{} enables the contract to safeguard their previously assigned funds from possible misappropriation in the submitted commitment. %

%% file: ProtocolSpecification.tex
\section{Protocol Details}
\label{sec:main}
\subsection{Smart-Contract State}
\label{subsec:sysmod}

\begin{table*}[]
\begin{tabular}{|c|p{15cm}|}
\hline
\textbf{Name} & \textbf{Description} \\ \hline
$pk_\omega$ & Public key of the operator \\ \hline
$apk$ & Aggregate public key of \vendors{} \\ \hline
$apk_a$ & Aggregate public key of \vendors{} whose signature was included in the last notarization \\ \hline
$g$ & Time that the last commitment generation event was to be triggered by the operator\\ \hline
$s$ & Time that the last notarized commitment was submitted \\ \hline
$R(\mathcal{M}(\mathbb{K}_t(-1)))$ & Merkle root of the last notarized commitment as of $t$\\ \hline
$\mathbb{X}$ & Pub keys of \vendors{} who exited/unregistered after the last block was notarized \\ \hline
$\mathbb{B}$ & Pub keys of \vendors{} who registered after the last block was notarized \\ \hline
$\mathbb{W}$ & Pub keys of \vendors{} who withdrew their funds after the last block was notarized \\ \hline
$\mathbb{M}$ & Pub keys of registered \vendors{} whose signatures were not included in the last notarized commitment and the number of consecutive commitments that each of these \vendors{} has missed signing so far%
\\ \hline %
$\mathbb{N}$ & Tracks the amount of funds \vendors{} in $\mathbb{M}(-x)$ have been assigned from $pk_c$, $\forall c$ whom $p\in \mathbb{M}(-x)$ have been assigned funds from in the last notarized block $\mathbb{K}_t(-x+1)$ that had their signature, $\forall x \in [1,\eta]$\\ \hline
$\mathbb{L}$ & Hash of payment transactions assigned to missing providers ($p \in \mathbb{M}$) in the $\kappa_p$  \\ \hline %
\end{tabular}
\caption{State of the \name{} smart-contract, representing the information it tracks}
\label{tab:state}
\end{table*}

The \name{} smart-contract tracks a minimal amount of information, as specified in Table~\ref{tab:state}.  Note that we refer to list elements by their indices depending on usage. %
At any time $t$, the contract stores two aggregate public keys of \vendors{} and only the last notarized block's Merkle root $R(\mathcal{M}(\mathbb{K}_t(-1)))$. %
Further \vendors{} in $\mathbb{X} \cup \mathbb{W} \cup \mathbb{B}$ have zero confirmed funds and those in $\mathbb{W} \cup \mathbb{B}$ wait until the next notarization to acquire new payments as confirmed funds, if any. %

The public keys of registered \vendors{} whose signatures were missing from the last notarized commitment are saved in $\mathbb{M}$. The contract also tracks the number of notarized commitments that these \vendors{} have consecutively missed. We use $\mathbb{M}(-x)$ to refer to \vendors{} whose signatures have been absent since the last $x = 1,2,\ldots$ commitments. Let $\eta$ denote the maximum value of $x$ (i.e. the maximum number of consecutive commitments missed by a non-signing \vendor{} in $\mathbb{M}$); note that $\mathbb{M} = \bigcup_{x=1}^{x=\eta} \mathbb{M}(-x)$. Note that the last notarized block $\kappa_p$ that was signed by a missing \vendor{} $p\in\mathbb{M}(-x)$ is simply $\mathbb{K}_t(-x+1)$ by definition of $\mathbb{M}(-x)$. The contract also tracks the public keys of consumers whom these missing \vendors{} were assigned funds from in the last notarized commitment that they signed (i.e. $\mathbb{Y}_{-x} = \{ s(T'),  \forall T' \in \mathbb{T}_p(\mathbb{K}_t(-x+1)), \forall p \in \mathbb{M}(-x)\}$)
, as well as the total amount that they had been assigned from each consumer (i.e. $\mu_{-x}^{pk_c} = \sum_{p \in \mathbb{M}(-x)} \sum_{T' \in \mathbb{T}_p(\mathbb{K}_t(-x+1))} \mathbbm{1}_{s(T') = pk_c} \mu(T')$). %
We use $\mathbb{N}$ to denote the resulting set of balances $\{\mu_{-x}^{pk_c}, \forall pk_c \in \mathbb{Y}_{-x}, \forall x \in [1,\eta]\}$, and use $\mathbb{N}(-x)^{pk_c}$ to refer to $\mu_{-x}^{pk_c}$. %
Appendix~\ref{app:MNExample} illustrates an example usage of $\mathbb{M}$ and $\mathbb{N}$. For each \vendor{} $p \in \mathbb{M}$, the contract also stores in list $\mathbb{L}$ the leaf node $L_p(\kappa_p)$ %
assigned to her in the last notarized block that had their signature. Note that the smart-contract's state is also accessible to anyone traversing the blockchain, and updates to these state values can be broadcast by the contract as well.

For ease of protocol description, we assume that consumer information (i.e. $pk_c$, $D_{c,t}$, $w_{c,t}^*$ for all $c$) is part of the smart-contract's stored state. However, this information is only required when consumers deposit additional funds into the contract or \vendors{} initiate withdrawal of assigned funds. These events are considerably infrequent in comparison with the periodic notarization events. %
This information can therefore be moved off-chain and instead represented just by a hash, as described in Appendix~\ref{sec:storageOptimization}. In exchange, marginally more computational work is expended when consumer information is required (during consumer top-ups or \vendor{} withdrawals). %

\subsection{Detailed Protocol Specification}
\label{sec:modules}
We now explain the core \name{} modules (cf. Figure~\ref{fig:flowchart}) in detail and illustrate how they fulfil the security properties identified in Section~\ref{subsec:secprop} (proofs in Appendix~\ref{sec:proofs}). We omit the specification of state updates for variables in Table~\ref{tab:state} that are straightforward. %
For instance, when a \vendor{} $p$ successfully registers with the \name{} smart-contract, we do not explicitly state the addition of $p$ to $\mathbb{B}$. $\mathbb{B}$ tracks this information by definition. %
We ensure that the \name{} contract is provided sufficient information during registration, notarization and withdrawal processing to ensure that updates to these variables can be correctly executed.%

\textbf{\vendorCap{} Registration}
To register, \vendor{} $p$ first sends a Proof of Possession (PoP) of her credentials to the operator. In other words, the \vendor{} uses her secret key $sk_p$ to sign her public key, generating $\sigma_{p, \text{init}} = \mathcal{S}(pk_p, sk_p)$. If the operator successfully verifies $p$'s signature on the PoP, i.e. $\mathcal{V}(pk_{p}, pk_p, \sigma_{p, \text{init}}) =1$, it signs the tuple of $p$'s public key and current timestamp $\tau_r$, i.e., $\omega$ generates $\sigma_{\omega, p} = \mathcal{S}(m$=$(pk_p,\tau), sk_{\omega})$. The operator returns $\sigma_{\omega, p}$ to the \vendor{}, who provides it along with $\tau_r$ to the smart-contract's enrollment function. 
\begin{algorithm}
    \SetKwInOut{Input}{Input}
    \SetKwInOut{Output}{Output}
    \DontPrintSemicolon
    \Input{$\sigma_{\omega, p}$, $\tau_r$, verified public-key of the caller $pk_p$}
    \Output{0 or 1}
    \KwData{$s$, $g$, $\mathbb{B}$, $\mathbb{X}$, $\mathbb{M}$, $\mathbb{W}$, $apk$, current time $t$}
    \If{not ($g + \gamma  + \delta < t < g + \beta - \delta$ and $s < \tau_r$) }
    {
        \KwRet $0$\;
    }
    \If{$pk_p \in \mathbb{B} \cup \mathbb{X} \cup \mathbb{M} \cup \mathbb{W}$} %
    {
        \KwRet $0$\;
    }
    \If{$\mathcal{V}(pk_{\omega}, m$=$(pk_p, \tau_r), \sigma_{\omega,p}) = 1$}
    {
        $apk$ = $\bot$ ? $apk= pk_p$ : $apk = apk\cdot{}pk_p$ \;
        \KwRet $1$\;
    }
    \caption{\vendorCap{} registration by Contract}
    \label{alg:contractProcessingRegistration}
\end{algorithm}
Algorithm~\ref{alg:contractProcessingRegistration} shows the contract's registration processing function; the \textbf{Data} field in the Algorithm denotes relevant internal state of the executing entity (the contract in this case). Let $t$ %
denote the time when the contract receives the registration request (note $t>\tau_r$). %
\vendorCap{} registrations are processed only when no freeze windows are currently active (cf. Figure~\ref{fig:plasmaTiming} in Section~\ref{sec:timing}) and no commitment has been notarized since the time reflected in the provided timestamp (Lines 1-3). %
As long $p$ is not known to have already registered (i.e. $p \notin \mathbb{B} \cup \mathbb{M} \cup \mathbb{W}$), or deregistered only since the last notarization (i.e. $p \notin \mathbb{X}$),
and the provided signature on $pk_p$ and $\tau_r$ is valid, the registration is successful and the contract updates the aggregated public key to include $p$'s key  (Lines 3-7). %
The operator $\omega$ must also register using this function to assign any portion of consumer payments that it retains as a fee to itself; it is treated like any other \vendor{} with respect to the notarization and withdrawal of its funds. Note that \name{} does not rely on the root-chain to support BLS account keys though this specification assumes for readability; Appendix~\ref{app:nonBLSaccount} explains this further.
\begin{lem}
\label{lem:useOwnKey}
$p$ can register only once unless $\omega$ colludes with $p$.
\end{lem}
Even if a colluding operator generates a $\sigma_{\omega,p}$ to allow an already registered \vendor{} to maliciously re-register%
, the resulting corruption to $apk$ does not affect \textit{confirmed funds} of any participant, as we later show.

\begin{algorithm}
    \SetKwInOut{Input}{Input}
    \SetKwInOut{Output}{Output}
    \DontPrintSemicolon
    \SetKwFunction{FgetSourceTransaction}{getSourceTransaction}
    \Input{$\kappa$, $\mathbb{C}_t$, $\tau$}
    \Output{$\sigma_{p,\kappa}$ or $\bot$}
    \KwData{$\mathbb{T}_p(\kappa_p)$, $\mathbb{M}$, $\mathbb{T}_m(\kappa_m) \forall m\in\mathbb{M}$, $\mathbb{W}$, $g$,$D_{c, t}$ for each $c$, $w_{c,t}^*$ for each $c$, current time $t$, hasWithdrawn=$\{0,1\}$}
    \If{not $g < \tau \le t < g + \gamma$}
    {
        \KwRet $\bot$\;
    }
    consPay, consOpBal= $\{\}$\;
    \For{$T'=(\mu', pk_p, pk_c)) \in \mathbb{T}_p(\kappa)$}
    {
        \If{hasWithdrawn=$0$}
        {
            $T'' = $getTransaction$(pk_c, \mathbb{T}_p(\kappa_p))$ \Comment{Cf.  Alg~\ref{alg:opComputingTree}} \;
            \If{$T'' != \bot$ and $\mu(T'') > \mu'$}
            {
                \KwRet $\bot$\;
            }
        }
        $T=$getSourceTransaction$(pk_c, \mathbb{C}_t)$\;
       \If{$T =\bot$}
        {
            \KwRet $\bot$\;
        }
        consOpBal$[pk_c] = \mu(T)$ \;
        consPay$[pk_c] = \mu'$ \;
    }
    regMissing\vendorsCap{} = $\mathbb{M} - \mathbb{W}$ \;
    \If{
    not [regMissing\vendorsCap{} $ \subseteq \mathbb{P}(\kappa)$ and $(\mathbb{T}_m(\kappa_m) \subseteq \mathbb{T}_m(\kappa), \forall m \in $regMissing\vendorsCap{}$)$]}
    {
        \KwRet $\bot$\;
    }
    \For{$T'=(\mu', pk_p, pk_c)) \in (\mathbb{T}(\kappa) \setminus \mathbb{T}_p(\kappa)).$flatten}
    {
       $T=$getSourceTransaction$(pk_c, \mathbb{C}_t)$\;
       \If{$s(T) \in $ consPay.keys} 
       {
            consPay$[pk_c]\pluseq\mu(T)$\;
       }
    }
    \If{consPay$[pk_c] > $ consOpBal$[pk_c]-w_{c, t}^*$}
    {
        \KwRet $\bot$\;
    }
    $\mathcal{M}'$ = merklize$(\mathbb{T}(\kappa))$ \;
    \eIf{$\mathcal{M}' = \mathcal{M}(\kappa)$}
    {
        $\sigma_{p,\kappa} = \mathcal{S}(m=(R(\mathcal{M}(\kappa)),\tau), sk_p)$\;
        \KwRet $\sigma_{p,\kappa}$\;
    }
    {
       \KwRet $\bot$\;
    }
    \SetKwProg{Fn}{Def}{:}{}
    \Fn{\FgetSourceTransaction{$pk_c$, $\mathbb{C}_t$}}{ 
        \For{($T,\sigma) \in \mathbb{C}_t$}
        {
            \If{$s(T) = pk_c$ and $\mathcal{V}(pk_c, T, \sigma) = 1$ and $\mu(T) \le D_{c,t}$}
            {
                \KwRet $T$\;
            }
        }
        \KwRet $\bot$\;
    }
    \caption{\footnotesize{Commitment} \small{Signing by registered \vendor{}}}
    \label{alg:providerSigningCommitment}
\end{algorithm}
\textbf{Commitment Generation } 
The block generation process executed by the operator every $\beta$ timesteps follows the description in Section~\ref{subsec:sysmod} and is shown in Algorithm~\ref{alg:opComputingTree} in Appendix~\ref{app:opAlg}.
After generating $\kappa$, the operator broadcasts $\kappa$, $\mathbb{C}_t$ and the current timestamp $\tau$ to \vendors{}. %
Algorithm~\ref{alg:providerSigningCommitment} details the process used by \vendors{} for verifying the validity of $\kappa$ and signing it. 
Note that $\mathbb{T}_m(\kappa_m)$ for all $m \in \mathbb{M}$ is known to all \vendors{} (though not stored by the contract) since it is broadcast by the contract's commitment verification module (explained later). 
The  binary flag hasWithdrawn is $1$ if the verifying \vendor{} $p$ withdrew their funds after the last notarization, i.e. $p \in \mathbb{W}$. %
$p$ verifiess that a commitment window is active (Lines 1-3), and that her confirmed funds does not decrease in this block as long as $p \notin \mathbb{W}$ (Line 4-8). 
The \vendor{} further ensures that a valid \textit{source transaction}, i.e. an off-chain payment from the consumer to the operator, accompanies each operator-generated payment transaction (Line 9-13). Finally, the \vendor{} guards herself against \textit{double-spend attacks} from the operator by verifying that, for each consumer who is listed as the source for a stated payment to $p$, the sum of payments promised to other \vendors{} with this consumer as the source does not exceed the operator-owned balance in the consumer's virtual channel (Lines 14-22). %
If these checks succeed and $\mathcal{M}(\kappa)$ is correctly generated from $\mathbb{T}(\kappa)$ (Lines 23-28), $p$ signs the tuple $(R(\mathcal{M}(\kappa)),\tau)$ and returns it.

Let set $\mathbb{A}_{t}$ and $\mathbb{M}_t$ respectively denote the public keys (or corresponding indices, based on usage) of registered \vendors{} who return the signed commitment to the operator within timeout duration $\gamma' < \gamma$ and those who do not. Here, $\gamma'$ is set by the operator such that $\gamma-\gamma'$ is sufficient duration for the operator to perform the remaining steps and submit the commitment to the root-chain. Let set $\mathbb{X}_t$ denote the public keys of previously-registered \vendors{} who had deregistered (i.e. exited) since the last block was notarized (which may be different from $\mathbb{X}$ tracked by the contract, as we see later). %
After verifying received signatures, the operator computes an \textit{aggregated root signature} $ars_{\kappa} = \prod_{p \in \mathbb{A}_t}  \sigma_{p, \kappa}$, and submits the following commitment to contract for notarization of $\kappa$:
$R(\mathcal{M}(\kappa))$, $\tau$, $\mathbb{X}_t$, and information on signing and missing \vendors{}. The \emph{signing \vendor{} information} consists of their aggregated public key $apk_{\text{active}}=\prod_{p \in \mathbb{A}_t} pk_p$ and $ars_{\kappa}$. Further, for signing \vendors{} whose signatures were missing in the last notarized commitment, i.e. $p \in \mathbb{M}$ and $p \in \mathbb{A}_t$, the operator provides $\mathbb{T}_p(\kappa_p)$. 
The \emph{missing \vendor{} information} consists of $\mathbb{M}_t$; for each missing \vendor{} who signed the previous notarized block, i.e. $p\in\mathbb{M}_t |  \kappa_p = \mathbb{K}_t(-1)$, the operator also includes: $\sigma_{p,  \text{init}}$ (collected during registration), $\mathbb{T}_p(\kappa)$, $P_p(\mathcal{M}(\kappa))$, 
$\mathbb{T}_p(\mathbb{K}_t(-1))$, and $P_p(\mathcal{M}(\mathbb{K}_t(-1))$. 
If the commitment is not submitted within $\gamma$, the operator must wait for the next commitment generation event. %
\begin{algorithm}
    \SetKwInOut{Input}{Input}
    \SetKwInOut{Output}{Output}
    \DontPrintSemicolon

    \Input{$R(\mathcal{M}(\kappa))$, $\tau$, $\mathbb{X}_t$, $apk_{\text{active}}$, $ars_{\kappa}$, $\mathbb{M}_t$, $(\sigma_{p,\text{init}}$, $\mathbb{T}_p(\kappa)$, $P_p(\mathcal{M}(\kappa))$, $\mathbb{T}_p(\mathbb{K}_t(-1))$, $P_p(\mathcal{M}(\mathbb{K}_t(-1))))$ for $p \in \mathbb{M}_t | \kappa_p = \mathbb{K}_t(-1)$, $\mathbb{T}_p(\kappa_p)$ for ($p \in \mathbb{A}_t$ and $p \in \mathbb{M}$)}
    \Output{1 or 0}
    \KwData{$apk$, $apk_a$, $\mathbb{M}$, $\mathbb{N}$, $\mathbb{X}$, $\mathbb{L}$, $\mathbb{B}$, $\mathbb{W}$, $R(\mathcal{M}(\mathbb{K}_t(-1)))$, current time $t$}
    \If{not $g_t' < s_t \le t < g_t' + \gamma$}
    {
        \KwRet 0 \;
    }
    \For{$p \in \mathbb{M}$ and $p \notin \mathbb{M}_t \cup \mathbb{X}_t$}
    {
        \Comment{H is the hash func. used by Merklize}\;
        \If{not $H(\mathbb{T}_p(\kappa_p)) = \mathbb{L}_p$}
        {
            \KwRet 0 \;
        }
    }
    \If {$\mathbb{X} \cap \mathbb{M}_t \neq \{\}$}
    {
        \KwRet 0 \;
    }
    \If{$\mathcal{V}(apk, m=({R}(\mathcal{M}(\kappa)), \tau), ars) = 1$}
    {
        \KwRet 1 \;
    }
    \eIf{$|\mathbb{X}_t| > 0$ or $\mathbb{M} \neq \mathbb{M}_t$}
    {
        \If{$apk_{\text{active}}\cdot{}\prod_{v \in \mathbb{M}_t} v\cdot{}\prod_{p \in \mathbb{X}_t} p \neq apk$}
        {
            \KwRet 0 \;
        }
    }
    {
        \If{$apk_{\text{active}} \neq apk_a$}
        {
            \KwRet 0 \;
        }
    }
    
    \If {
    not $[\mathbb{X}_t \subseteq \mathbb{X}$ and $\mathcal{V}(apk_{\text{active}}, (R(\mathcal{M}(\kappa)), \tau), ars_\kappa)=1]$}
    {
         \KwRet 0 \;
    }
    \For{$p \in \mathbb{M}_t - \mathbb{M} - \mathbb{B} - \mathbb{W}$}
    {
        \If{not $\mathcal{V}(pk_p, pk_p, \sigma_{p,\text{init}}) = 1$} 
        {
             \KwRet 0 \;
        }
         \Comment{checkMP verifies a Merkle Proof}\;%
        \If{
        not $[$checkMP$(P_p(\mathcal{M}(\kappa)), R(\mathcal{M}(\kappa)))$ and checkMP$(P_p(\mathcal{M}(\kappa)), R(\mathcal{M}(\mathbb{K}_t(-1))))]$}
        {
            \KwRet 0 \;
        }
        \If{not $(\mathbb{T}_p(\kappa) \supseteq \mathbb{T}_p(\mathbb{K}_t(-1)))$}
        {
            \KwRet 0 \;
        }
    }
    \If{$|\mathbb{X}_t| > 0$}
    {
        $apk = apk_{\text{active}}\cdot{}\prod_{p\in\mathbb{M}_t}p$ \;
    }
    \KwRet 1 \;
    \caption{Commitment Verification}
    \label{alg:commitmentVerification}
\end{algorithm}

\textbf{Commitment Verification} 
When the operator submits a new block commitment at $t$, the smart-contract performs the verification steps described in Algorithm~\ref{alg:commitmentVerification}, where return values of $0$ and $1$ indicate commitment rejection and acceptance respectively. %
For \vendors{} whose signatures were not included in the last notarized block but included in the current one, the contract requires the notarized payment transactions that they had last signed for (Lines 3-6). Since these \vendors{} must be removed from $\mathbb{M}$, $\mathbb{N}$ and $\mathbb{L}$, this provides the contract with necessary information to correctly update the state.
If all registered \vendors{} have signed the tuple of the provided Merkle root and timestamp $\tau$, the verification immediately succeeds (Lines 9-10) since their signature conveys that their double-spend and safety checks on the generated commitment succeeded (cf.  Algorithm~\ref{alg:providerSigningCommitment}).
If, however, only a subset of registered \vendors{} have signed the commitment, then extra steps (Lines 11-26) are needed to ensure \vendor{} safety and data availability for those whose signatures are not included.

First, \vendors{} who recently exited must be removed from $apk$. Though %
the contract hence stores their public keys in $\mathbb{X}$, the contract cannot determine which exited \vendors{} were registered. Since the contract neither tracks individual keys of registered \vendors{} nor validates the full Merkle tree corresponding to a committed root, the operator may well assign payments even to unregistered \vendors{} in a generated Merkle tree (they are now guaranteed security of their notarized funds or protected from data availability attacks). The contract must update $apk$ to remove \vendors{} who have exited while retaining registered \vendors{}; however, requires knowing 1) which exited \vendors{} were registered, and 2) the public keys of remaining \vendors{} to recompute $apk$. Second, registered (non-exited) \vendors{} who have not signed the submitted commitment must be identified so that their previously assigned funds can be secured against any malfeasance in this commitment.This is challenging for similar reasons; the individual keys of \vendors{} are not stored. Algorithm~\ref{alg:commitmentVerification} is designed to efficiently overcome this problem. 

The contract first verifies that $apk$ matches the key generated from aggregating the provided keys of active, missing, and exited \vendors{} (Lines 11-16). %
Then, the contract checks the validity of $\mathbb{X}_t$ and the provided $ars_\kappa$ against the provided $apk_{\text{active}}$ (Lines 17-18).
At this point, it is not yet guaranteed that all registered \vendors{} have been accounted for in $ars_{\kappa}$; potential attacks against these checks are demonstrated in the proof of subsequent Theorem~\ref{thm:trackingMissingProviders}. %
Additional checks are hence performed. If missing \vendor{} $p$ is in  $\mathbb{M}$
, then $p$'s credentials have already been verified through the notarization process for a past commitment and its confirmed funds already secure. If missing \vendor{} $p  \in \mathbb{B} + \mathbb{W}$, then $p$'s credentials have already been verified though a recent registration or withdrawal event in the root-chain and she has zero confirmed funds to secure as she joined only after the last block was notarized or withdrew all her funds since. For the rest of the missing \vendors{}, the contract checks that a correct PoP has been provided for each, that the provided Merkle proofs are correct, and that the payment transactions assigned to them in the current commitment is at least equivalent to the transactions assigned in the previous commitment (Lines 19-26). In that case, the verification of the submitted commitment succeeds. Further $apk$ is updated to remove deregistered \vendors{} (Line 27-28).
Through these checks, the contract can assess whether all registered \vendors{} have been accounted for in the provided commitment \textit{despite no explicit long-term record of their public keys, balances or payment transactions}. %
If the commitment is accepted, the contract's state variables (in Table~\ref{tab:state}) are updated as necessary. Note that all non-signing \vendors{} of this commitment are identified are tracked appropriately; if a newly missing \vendor{} had been in $\mathbb{B}$ or $\mathbb{W}$, the contract simply stores the $\mathbb{L}_p=H(\{0\})$ for her. The contract broadcasts any updates to $apk$ as well as any additions and deletions of \vendors{} $p$ to $\mathbb{M}$ and their corresponding $\mathbb{T}_p(\kappa_p)$. 
\begin{thm}
\label{thm:trackingMissingProviders}
Suppose $p$ is a registered \vendor{} whose signature is not included in $ars_{\kappa}$. If the commitment is accepted, $p$ was detected as a non-signing \vendor{}, i.e.$pk_p \in \mathbb{M}_t$.
\end{thm}

\begin{thm}
\label{thm:excludingNonRegisteredProviders}
If a commitment is accepted, $apk_{\text{active}}$ and $\mathbb{M}_t$ only consist of \vendors{} who had registered with the \name{} smart-contract at some earlier time, and $\mathbb{X}_t$ only consists of \vendors{} who had registered at some earlier time and deregistered after the last notarization. 
\end{thm}

Note that this leaves room for 1) a registered \vendor{} to be simultaneously present in $apk_{\text{active}}$ and $\mathbb{M}_t$ of an accepted commitment, and 2) for a deregistered \vendor{} in $\mathbb{X}_t$ to be simultaneously present in $apk_{\text{active}}$, and 2) for a deregistered \vendor{} who had deregistered in earlier notarizations (i.e. not tracked in $\mathbb{X}$) to be present in $apk_{\text{active}}$ and/or $\mathbb{M}_t$ of subsequently accepted commitments. These states only occur when a \vendor{} has maliciously registered with the \name{} smart-contract despite being already registered (by colluding with the operator, cf. Lemma~\ref{lem:useOwnKey}). However, Lines 7-8 in Algorithm~\ref{alg:commitmentVerification} combined with the construction of our withdrawal module ensures that these apparent corruptions are ineffective in violating \textit{Consumer and \vendorCap{} Safety}.

\begin{algorithm}
    \SetKwInOut{Input}{Input}
    \SetKwInOut{Output}{Output}
    \DontPrintSemicolon

    \Input{$\mathbb{T}_p$, $\mathbb{C}_p'$, (optional) $\mathcal{P}_p(\mathcal{M})$ %
    }
    \Output{Amount of \currency{} to transfer to $p$}
    \KwData{$apk$, $\mathbb{M}$, $\mathbb{N}$, $\mathbb{X}$, $\mathbb{B}$, $\mathbb{W}$,$\mathbb{L}$, ${R}(\mathcal{M}(\mathbb{K}_t(-1)))$, $D_{c,t}$ for consumers $c$, $\mu'_{c,t}$, %
    $w_{c,t}^*$,  %
    $g_t'$, current time $t$}
    fundsToTransferToP $\gets 0$ \;
    blocksPMissed $\gets 0$ \;
    \If{not $g_t' + \gamma + \delta < t < g_t' + t_a - \delta$}
    {
        \KwRet 0 \;
    }
    \If{$p \in \mathbb{X} \cup \mathbb{B} \cup \mathbb{W}$}
    {
        \KwRet 0 \;
    }
    \If{$p \in \mathbb{M}$}
    {
        \If{$H(\mathbb{T}_p) != \mathbb{L}_p$}
        {
            \KwRet 0 \;
        }
        blocksPMissed $\gets x | p \in \mathbb{M}(-x)$\;
    }
    \If{$p \notin \mathbb{M}$ and not checkMP$(\mathcal{P}_p(\mathcal{M})), R(\mathcal{M}(\mathbb{K}_t(-1)))$}
    {
        \KwRet 0 \;
    }
    \For {$T'=(\mu', pk_p, pk_c) \in \mathbb{T}_p$}
    {
        $T$ = getSourceTransaction$(pk_c, \mathbb{C}_p')$ \Comment{Cf.  Alg~\ref{alg:providerSigningCommitment}}\;
        \If{$T = \bot$}
        {
            \KwRet 0 \;
        }
        \If{$\mu(T) > \mu'_{c,t}$}
        {
            $ \mu'_{c,t} = \mu(T)$\;
        }
        fundsToTransferToP $\pluseq \text{min}\{\text{max}\{\mu_{c,t}' - \sum_{i = \eta}^{\text{blocksPMissed}-1} \mathbb{N}(-i)^{pk_c} - w_{c,t}^*, 0\}, \mu'\}$
    }
    \KwRet fundsToTransferToP \;
    \caption{Contract processing $p$'s withdrawal}
    \label{alg:withdrawal}
\end{algorithm}

\textbf{Withdrawal}
Algorithm~\ref{alg:withdrawal} describes the procedure used by the contract's withdrawal function to determine the amount of funds to be transferred to a \vendor{} (or operator) when invoked.
As input, \vendor{} $p$ submits a set of transactions $\mathbb{T}_p$ that assigns payments to her, a set $\mathbb{C}_p'$ of off-chain transactions between each consumer whose funds $p$ has been assigned payments from and the operator. Note that $\mathbb{C}_p' \subseteq \mathbb{C}_{t'}$, where $\mathbb{C}_{t'}$ denotes the set of off-chain transaction from each consumer to the operator that was revealed to $p$ as a part of some commitment-generation process at time $t'$. %
The \vendor{} must also specify whether the withdrawal is a permanent exit (i.e. deregistration). If $p$'s signature was included in $\mathbb{K}_t(-1)$, $p$ also submits a corresponding Merkle proof for $\mathbb{T}_p$.

Note that a \vendor{} who has already withdrawn funds since the last block was notarized (or has registered only since) cannot maliciously initiate a withdrawal %
since \textit{all of the \vendor{}'s confirmed funds} %
are transferred to her upon a successful withdrawal (Lines 5-6). %
If the \vendor{}'s signature has not been included in the last notarized block, the function expects that the provided transaction set $\mathbb{T}_p$ is equivalent to what she last signed for (i.e. $\mathbb{T}_p(\kappa_p)$) (Lines 7-10). On the other hand, if the \vendor{}'s signature was included in the last notarization, then the contract requires a valid Merkle proof for $\mathbb{T}_p$ against $R(\mathcal{M}(\mathbb{K}_t(-1)))$ (Lines 11-12). %

The contract then iterates over each payment transaction $T' \in \mathbb{T}_p$ to determine the total funds to be transferred to $p$ (Lines 13-20). First, each $T'$ must be associated with a valid consumer$\rightarrow$operator transaction and associated signature $(T,\sigma)$ in the provided input $\mathbb{C}_p'$. The contract then ensures that no more funds are withdrawn against the source consumer $s(T')$ than what the consumer has assigned as off-chain funds to the operator. $\mu(T') < \mu(T)$ does not suffice as prior \vendor{} withdrawals may have been processed against this consumer $s(T)$.
While total withdrawals against a consumer's channel $w_{c,t}^\ast$ is known to the contract (since withdrawals happen through it), it may not know the most recent value of the operator-owned balance $\mu_{c,t}^\ast$ in that channel since consumer payments to the operator happen off-chain. We hence use $\mu_{c,t}'$ to represent the highest operator-owned balance in the channel with $c$ \textit{known to the contract} (as revealed by the consumer$\rightarrow$operator source transactions submitted by \vendors{} during withdrawals). Note $\mu_{c,t}' \le \mu_{c,t}^* \le D_{c,t}$. %
Finally, the contract must also secure funds of \vendors{} whose signatures have been missing in the last notarization, since double-spend attacks may have been launched against these \vendors{} in that block (by the operator and colluding \vendors{}). The state $\mathbb{N}$ stored by the contract that tracks missing \vendors{}' total funds against each consumer that they have stake in is used for this. %
After the withdrawal, the contract's state variables (in Table~\ref{tab:state}) are updated as necessary.

\begin{thm}
\label{thm:consumerFundsSafety}
\name{} ensures Consumer Safety.
\end{thm}

\begin{thm}
\label{thm:providerSafety}
\name{} ensures \vendorCap{} Safety.
\end{thm}

\begin{cor}
\label{cor:dataAvailability}
\name{} provides Data Availability and Income Certainty.%
\end{cor}

%% file: Eval.tex
\section{Evaluation}
\label{sec:evaluation}
We evaluate the computational and monetary costs incurred by \name{} for its main recurring operations, commitment generation by the operator and commitment verification by the smart-contract. We use ZK Rollup as the baseline in our analysis; ZK Rollup has become a popular solution for non-pairwise off-chain payments and has recently been deployed by multiple teams on Ethereum 2.0 mainnet~\cite{loopring, matterLabs}. Note that it also satisfies almost all properties identified in Table~\ref{tab:secComparison}. %

\textbf{Notation} Let $n$ be the number of payments made by consumers during $\beta$ and $p_u$ the number of unique payment recipients. Consider \name{} and ZK Rollup notarization executed at the end of $\beta$. %
Let $p_r$ and $c_r$ be the total number of registered \vendors{} and consumers, and $c_u$ the average number of unique consumers that a \vendor{} has been assigned payments from. Let $p_m$ , $p_{m'}$ and $p_a$ respectively be the number of non-signing \vendors{} in the submitted commitment, the subset of these that had not signed the previous notarization either (note $p_{m'}<p_m$), the number of signing \vendors{} who had not signed the previous commitment.  Let $p_x$, $p_b$ and $p_w$ respectively be the number of deregistered \vendors{} , newly registered \vendors{} and the number of registered \vendors{} who had withdrawn their funds since the last notarization. For the zkSNARK circuit used in the ZK Rollup, let $g'$ be the number of gates, $w'$ the number of wires and $l'$ the number of known circuit inputs (for maximum instance size). 

\begin{table}%
\centering
  \begin{tabular}{|l|l|l|l|l|} \hline
    \multirow{2}{*}{} &
      \multicolumn{2}{c|}{\textbf{Runtime} } &
      \multicolumn{2}{c|}{\textbf{\# Pairing \& Exp.}} \\
      & {PayPlace} & {ZKR} & {PayPlace} & {ZK Rollup}  \\ \hline
Op. & $O(p_r)$ & $O(n)$ & $2p_r$ & $\frac{n}{z_{\text{max}}} (4g'+w'-l')$ \\ \hline
Mer. & $O(c_u\cdot{}p_r)$ & $O(1)$ & $2c_u$ & $0$ \\ \hline
  \end{tabular}
  \caption{Comparing off-chain computational load and runtime.} %
  \label{tab:opVendorLoad}
\end{table}

\begin{figure}
    \centering
    \includegraphics[width=.9\columnwidth,trim={.6cm .6cm .6cm .6cm}, clip]{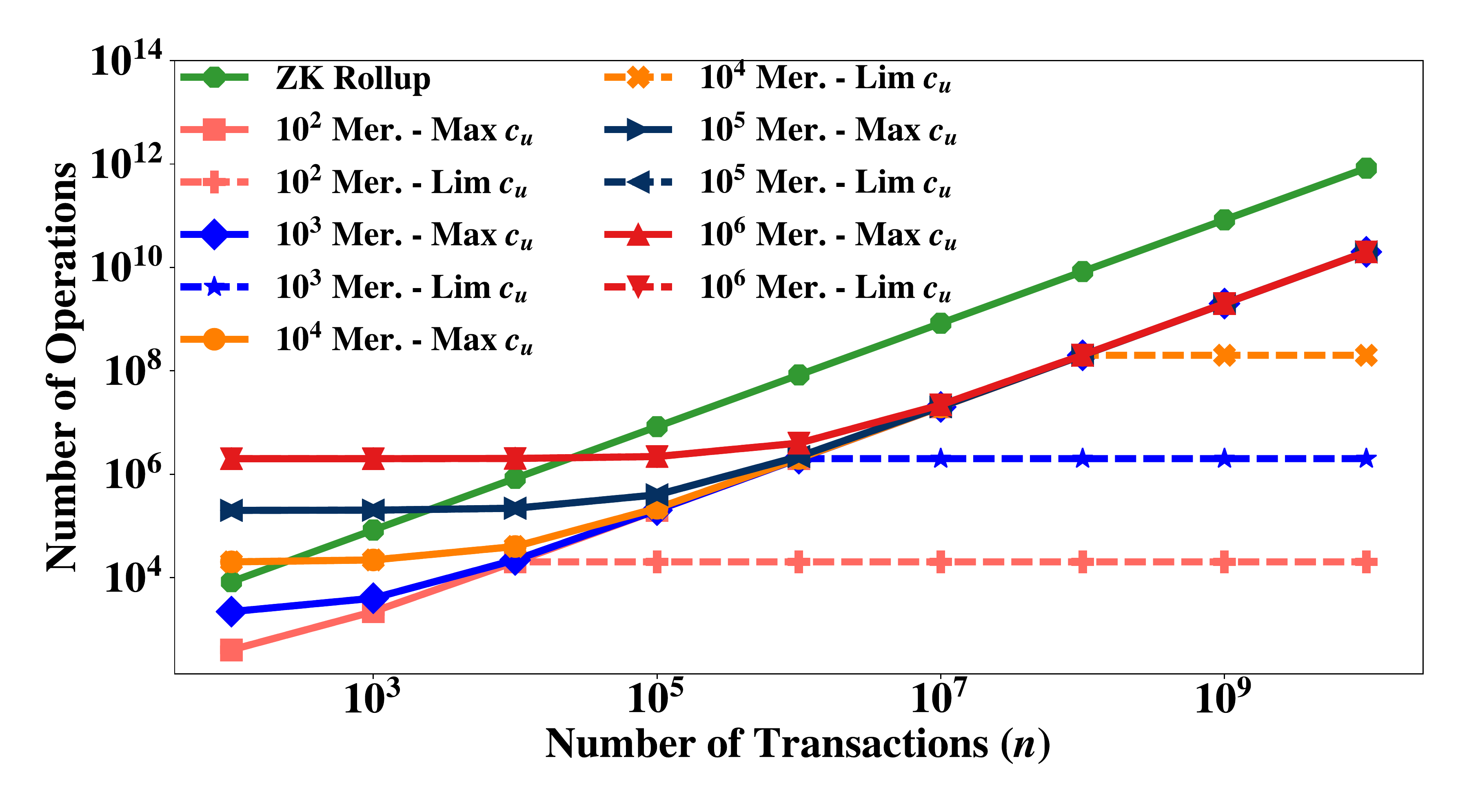}
    \caption{The number of pairings and exponentiations scales linearly with $n$ for ZK Rollup and linearly in $p_r+c_u$ for \name{}. As $n$ increases, \name{} incurs orders of magnitude lower computational load, even for large marketplaces (high $p_r$) with maximum $c_u$.}
    \label{fig:offchainComputationalLoad}
\end{figure}
\textbf{Off-Chain Computational Overhead} %
Computing zkSNARK proofs is highly expensive, with bilinear pairings and group exponentiations dominating all other involved operations in cost. We sidestep this in \name{} by offloading some computation to each \vendor{}, who protects her assigned funds in a block by the verification steps she performs before signing it. We hence assess the runtime complexity and dominant computational load for both the operator and \vendors{}. %
For ZK Rollup, we consider Groth16 SNARKs~\cite{groth2016size}, which are in wide use and recently deployed~\cite{loopringGroth16, zokratesGroth16, matter} in live ZK Rollup implementations. The number of off-chain payment transactions that can be included in the proof depends on the arithmetic circuit and further optimizations; let this maximum number of transactions be denoted by $z_{\text{max}}$. %
For $n$ total transactions then, $\lceil n/z_{\text{max}} \rceil$ prover computations need to be performed by the operator. Table~\ref{tab:opVendorLoad} reflects the corresponding amortized computational load and runtime complexity. 
For \name{}, note that Merkle root of a $p_r$-leaved tree can be computed in $O(log_2(p_r))$ time~\cite{szydlo2004merkle} (given $O(log_2(p_r)$ space). %
Hence, the operator's runtime complexity for commitment generation is dominated by the signature verification operations ($p_r$ verifications in the computational worst-case when all \vendors{} return signed commitments) %
while \vendors{}' by the double-spend verification checks they do before signing a block (Algorithm~\ref{alg:providerSigningCommitment}). %
In essence, computational costs for a ZK Rollup operator scales linearly in $n$ while for \name{} it is primarily a function of $p_r$ and $c_u$. For both \name{} and ZK Rollup, the operator and \vendor{} runtimes can be reduced to $O(1)$ with arbitrary space complexity (i.e. these computations are fully parallelizable).

We now empirically study the computational load of the two techniques in practice (in terms of the expensive cryptographic operations - pairings and exponentiations). 
We vary the number of transactions during $\beta$ from $100-10$B. To put this in perspective, Amazon is estimated to process roughly $27$M order per day and Uber roughly $15$M rides per day worldwide~\cite{uberRide1,uberRide2,amazonStat}. 
Recent data~\cite{matter, loopringZKR} from ZK Rollup systems indicate a capacity of $2-3$K transactions per proof, hence we conservatively set $r_\text{max}=3000$ and consider a load of only $150$K total pairings and exponentiations per SNARK proof computation (in practice, $150$K is approximately the number of constraints in the Rollup circuit reported by benchmarks, yielding much higher $4g' + w' - l'$).
As Figure~\ref{fig:offchainComputationalLoad} depicts (log-log scale), the dominant off-chain computational load in ZK Rollupincreases linearly in the number of transactions. 
\begin{table*}[]
\centering
\begin{tabular}{|l|p{1.5cm}|p{2.5cm}|p{5.5cm}| p{2.5cm}|}
\hline
\textbf{Operation} & \textbf{Best Case} &\textbf{Average Case} & \textbf{Worst Case} \\ \hline
Bilinear Pairings  & $O(1)$ & $O(1)$ & $O(p_m-p_{m'}-p_b-p_w)$\\ \hline %
Multiplications in $\mathbb{G}_1$ & $O(1)$ & $O(1)$ & $O(p_m + p_x)$\\ \hline
Hashing into $\mathbb{G}_0$ & $O(1)$ & $O(1)$ & $O(p_m-p_{m'}-p_b-p_w)$
\\ \hline
Non-$\mathbb{G}_0$ Hashes & $O(1)$ & $O(1)$ & $O(p_a +(p_m-p_{m'}-p_b-p_w)\cdot{}log_2(p_r))$\\ \hline
\end{tabular}
\caption{Best, Worst and Average-case. Runtime Complexity of notarization in \name{}, categorized by the operation type.}
\label{tab:timeComplexity}
\end{table*}
From Table~\ref{tab:opVendorLoad}, however, it is evident that the number of such operations is \name{} is not a function of $n$ but of $p_r$ and $c_u$. We hence vary $p_r$ from $100-1$M \vendors{}; to put this in perspective, Amazon and Uber have around $2$M sellers and drivers respectively. To assess worst-case load, we let $c_u$ equal the number of transactions per \vendor{}. For instance, if $n=1000$ and $p_r=100$, then we evenly distribute the orders across \vendors{} as $10$ orders per \vendor{}, and assume a worst-case scenario of $1$ unique consumer per order, resulting in $c_u=10$. If $p_r > n$, only $n$ \vendors{} receive orders ($c_u=1$ for them). In practice, marketplaces may involve recurrent transactions between consumers and \vendors{} (e.g. due to co-location and especially if $\beta$ spans longer time periods). %
We hence also consider $c_u$ no greater than $p_r$ to model this. We refer to this as the "Limited" or Lim $c_u$ in Figure~\ref{fig:offchainComputationalLoad} and the former as the "Maximum" or Max $c_u$. %
As we see from Figure~\ref{fig:offchainComputationalLoad}, even as the number of \vendors{} increase exponentially, the computational load across the operator and all \vendors{} in \name{} is orders of magnitude lower than ZK Rollup as the number of marketplace transactions increase exponentially. %
In-practice, the \name{} operator factors in for typical $p_r$ and $u_c$ in the marketplace to estimate the duration $\gamma$ required to execute these off-chain operations. %

\textbf{On-Chain Notarization Complexity} 
We next assess the on-chain runtime complexity of \name{} and ZK Rollup notarizations. %
For \name{}, we further categorize this by three scenarios and the core operations involved. We consider the best-case scenario as $p_x=0$, $p_a=0$, $p_m = 0$ (i.e. all registered \vendors{} have signed the submitted commitment with no one having recently exited or missed the previous one), and otherwise as the worst-case. We consider the case where all non-signing \vendors{} had missed the previous notarization as well (i.e. $p_{m'} = p_m$), all signing \vendors{} had signed the previous notarization as well (i.e. $p_a = 0$), and no \vendors{} exited since the last notarization (i.e. $p_x = 0$) as the average case. In practice, \vendors{} registering and exiting the system is likely infrequent in comparison with notarization events, especially since \name{} guarantees safety of their notarized funds and does not incur any exit games.

Table~\ref{tab:timeComplexity} specifies the runtime complexities for \name{}. \name{} is overwhelmingly $O(1)$ in the number of transactions $n$ and recipient providers $p_r$ (cf. Algorithm~\ref{alg:commitmentVerification}) except in the worse-case, where the hashes required for verifying Merkle proofs for newly non-signing \vendors{} scales logarithmically with $p_r$. This directly results in \textit{On-Chain Efficiency} in \name{}. In the best-case \textit{as well as} average-case scenario, \name{} is O(1) in all operations. This provides an important insight; the complexity of the smart-contract's notarization module in \name{} does not increase \textit{even for arbitrarily large quantities of non-signing \vendors{}} as long as these non-signing \vendors{} remain inactive for multiple notarizations once they become inactive. 
Further, well-known marketplace operators like Amazon and Uber are often atleast semi-trusted and \vendors{} may well opt to only intermittently participate in the signing process to receive additional income, leading often to low $p_m - p_{m'}$ and $p_a=0$. In the worst case, the most expensive operation, Bilinear Pairings, scales only in $p_m - p_{m'}$. %
For ZK Rollup, though the Groth16 SNARK verification can be run in constant time, only $z_\text{max}$ transactions can be included in one proof. Hence the amortized time complexity is $O(n)$.

\textbf{On-Chain Notarization Costs} To assess the on-chain computational resources required for frequent notarization operations, %
estimate the gas costs incurred in the Ethereum blockchain for ZK Rollup and \name{} notarizations. %
As of the Istanbul network update~\cite{istanbul}, the SNARK verification for Rollup is estimated to cost approximately $300$K gas~\cite{loopringZKR,zksync}. %
Transactions further have to be published on the root chain at least in CALLDATA to ensure data availability~\cite{zkrollup}. CALLDATA is then a recurring cost %
of $16$ gas~\cite{istanbul} per byte and each included transaction is $15$ bytes. We add an overhead of $50$K gas to account for additional costs, e.g., due to logging, storage slot modifications, etc. as done previously~\cite{zkrollup}. %
For \name{}, %
we set $p_a$, $p_b$ and $p_x$ to $0$ as these represent negligible overheads. As evident from Table~\ref{tab:timeComplexity}, the computational complexity is affected significantly by $p_m$, $p_{m'}$ and $p_r$; we study those here.
While native support for BLS12-381 curve operations (i.e. pre-compiles) is being planned in Ethereum~\cite{blscurvePrecompile,eth2, blspairingeth}, %
the alt\_bn128 curve is mainly used for zkSNARKs and BLS signatures. We hence use gas costs charged by the alt\_bn128 pre-compiled contract offered in Ethereum~\cite{altBn128Precompile} to estimate the cost for BLS signature operations.
Parings cost $34\text{K}\cdot{}\text{num}_{\text{pairings}} + 45$K  gas, the cost of a key multiplication in $\mathbb{G}_1$ (for public key multiplication) is $150$, and the cost per keccak256 hash is $42$. We estimate a higher cost of $100$ gas for hashing into $\mathbb{G}_0$. Note that verifying $x$ PoPs can be done with $x+1$ pairings (rather than $2x$)~\cite{boneh2003aggregate, boneh2018compact}. %
We assume a fixed overhead of $30$K (for addition/assignment operations, broadcasting events) and a variable overhead of $10$K in the number of additional non-signing \vendors{} in the submitted commitment (i.e. $p_m - p_{m'}$) to account for the storage and broadcast operations involved.

\begin{figure}[t!]
  \centering
  \begin{subfigure}[b]{\textwidth}
    \includegraphics[width=.9\textwidth,trim={1cm 1cm 1cm 1cm},clip]{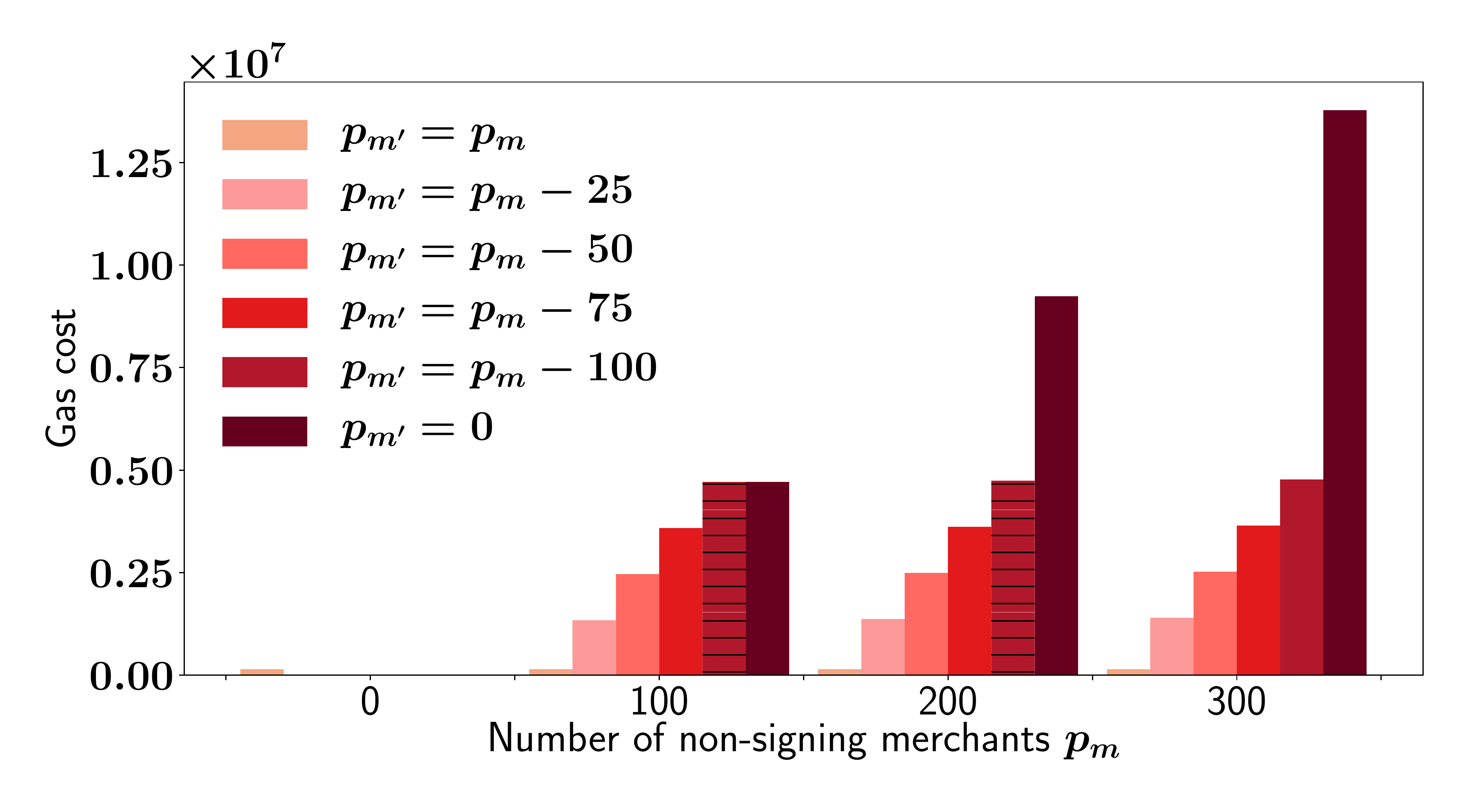}
    \caption{Gas cost for \name{} notarization primarily scales with the $p_m-p_{m'}$ rather than with $p_m$. It increases with \textit{additional} \vendors{} who have not signed the notarization compared to the previous one.
    \newline
    }
    \label{fig:missingProviderRelation}
  \end{subfigure}
  \begin{subfigure}[b]{\textwidth}
    \includegraphics[width=.9\textwidth,trim={1cm 1cm 1cm 1cm},clip]{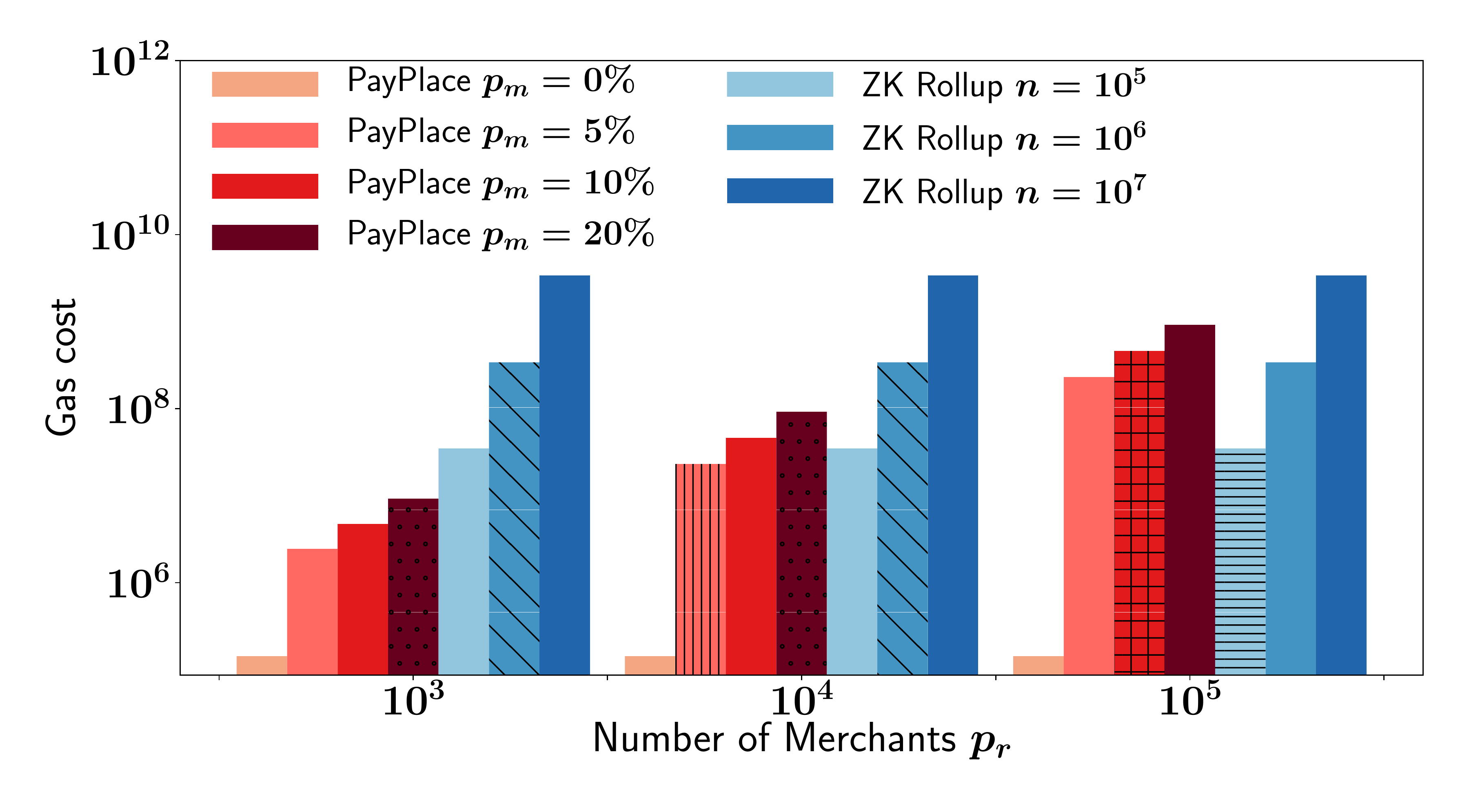}
    \caption{Gas cost for ZK Rollup increases with $n$ while worst-case \name{} is orders of magnitude cheaper when $p_m$ is relatively low wrt $n$.}
    \label{fig:zkrvspp}
  \end{subfigure}
    \caption{Estimated gas costs for (a) \name{} notarization with $p_r=1000$ as a function of $p_m$ and $p_{m'}$ and (b) Worst-case \name{} ($p_{m'} = 0)$) vs ZK Rollup notarizations as a function of $n$ and $p_m$.}
    \label{fig:onChainGas}
\end{figure}

Figure~\ref{fig:onChainGas} illustrates these estimated gas costs in \name{} and ZK Rollup notarizations. We set $p_r=1000$ in Figure~\ref{fig:missingProviderRelation}, and vary the value of $p_m - p_{m'}$ for different values of $p_m$. Crucially, we observe that even as the number of non-signing \vendors{} $p_m$ increases, \textit{the gas required is near-constant} as long as $p_{m}-p_{m'}$ is the same. This directly corroborates the analysis in Table~\ref{tab:timeComplexity}, where only multiplications in $\mathbb{G}_1$ scales with $p_m$ corresponding to Line 10 from Algorithm~\ref{alg:commitmentVerification}. Since bilinear pairings far exceed the rest in cost per operation and scales linearly only with $p_m - p_{m'}$, we see in Figure~\ref{fig:missingProviderRelation} that notarization costs predominantly increase only in the number of \textit{additional} non-signing \vendors{} in a commitment and not the \textit{recurring} non-signers. In-fact, very little gas is expended when $p_m = p_{m'}$ irrespective of the value of $p_{m}$. 
In the marketplace context, this implies that notarization costs are high only if \vendors{} tend to oscillate between being active and inactive during consecutive commitment submission windows (resulting frequently in $p_{m'} \gg p_m$) . In practice, we may expect \vendors{} to participate reliably in the notarization process to receive additional income on-time, or to frequently miss the signing process and only sporadically accrue income (i.e. the operator is highly trusted). %
Essentially, \name{} is designed so that \vendors{} impose a relatively high cost the first notarization that they miss after being active, but little cost for subsequent ones.
Figure~\ref{fig:zkrvspp} (log scale) compares gas estimates for ZK Rollup vs the worst-case gas estimates for \name{} (i.e. when $p_{m'}=0$) for different $p_r$ based on the fraction of the \vendor{} population that is non-signing (i.e. $p_m$). Unsurprisingly, the former scales linearly in $n$, as also seen in Table~\ref{tab:timeComplexity} and is $O(1)$ in $p_m$ and $p_r$; however, even with $p_{m'}=0$, the latter is often orders of magnitude lower in cost (even for large $p_m$ if $p_m \ll n$). 
It is evident that operators choosing between these two off-chain payment solutions must assess the expected transaction volume and the \vendor{} population. When \vendors{}' devices can reasonably be expected to participate in the signing process (once per day or hour, based on $\beta$)  or atleast stay in their active or inactive states for extended periods, \textit{\name{} is highly beneficial by scaling throughput at no marginal gas cost.} 

%% file: Discussion.tex
\section{Discussion}
\label{sec:discussion}

In practice, most blockchains impose limits on the amount of computations that can be performed as a part of a single transaction. %
The notarization process executed by the \name{} smart-contract, however, scales linearly with factors like $p_m-p_{m'}$ in the worst-case analysis from Table~\ref{tab:timeComplexity}, imposing limits on these factors to stay within the block limit. %
One way to overcome this is to allow operators to force the exit of such non-signing \vendors{} using the withdrawal module before submitting the notarization. %
Alternatively, the operator may use a zkSNARK to prove to the contract that funds previously assigned in the last notarization to newly non-signing \vendors{} $p_m - p_{m'}$ have been included in the Merkle tree whose root has been submitted for notarization. %
Finally, the operator may entirely avoid this linear cost of non-signing \vendors{} by instead submitting an additional aggregate signature of newly joined \vendors{} (since the last notarization) on the previous notarized root. This ensures that these new \vendors{} can detect if the operator has malicious omitted a \vendor{} who was assigned funds in the previous commitment from the current one; hence signing \vendors{} can still protect themselves from double-spend attacks.

%% file: Conclusion.tex
\section{Conclusion}
\label{sec:conclusion}
We develop \name{}, an off-chain payment protocol optimized for large marketplaces that overcomes liquidity and capital drawbacks of previous solutions while keep the root-chain footprint low. \name{} takes advantage of the presence of marketplace operators and introduces them as semi-custodial intermediaries in the payment process. Consumers pay the operator \textit{off-chain }during order placement, and the operator periodically forwards the accrued payments \textit{off-chain} without requiring any liquidity.
Our construction results in highly usability; consumers are able to view their off-chain payments to the operator as transactions in a unidirectional payment channel, while \vendors{} are guaranteed safety of their notarized funds even if they are arbitrarily offline.
We show that, based on how frequently \vendors{} oscillate between being available to sign notarizations and not, \name{} is potentially orders of magnitude cheaper in on-chain and off-chain execution costs compared to the state-of-the-art technique for non-pairwise off-chain payments, Zero Knowledge Rollups. %

%% file: Appendix.tex
\appendices

\section{Illustrating Merkle Tree generated by the \name{} Operator}
\label{app:merkleFigure}
\begin{figure}[!htp]
    \centering
    \includegraphics[width=.7\columnwidth]{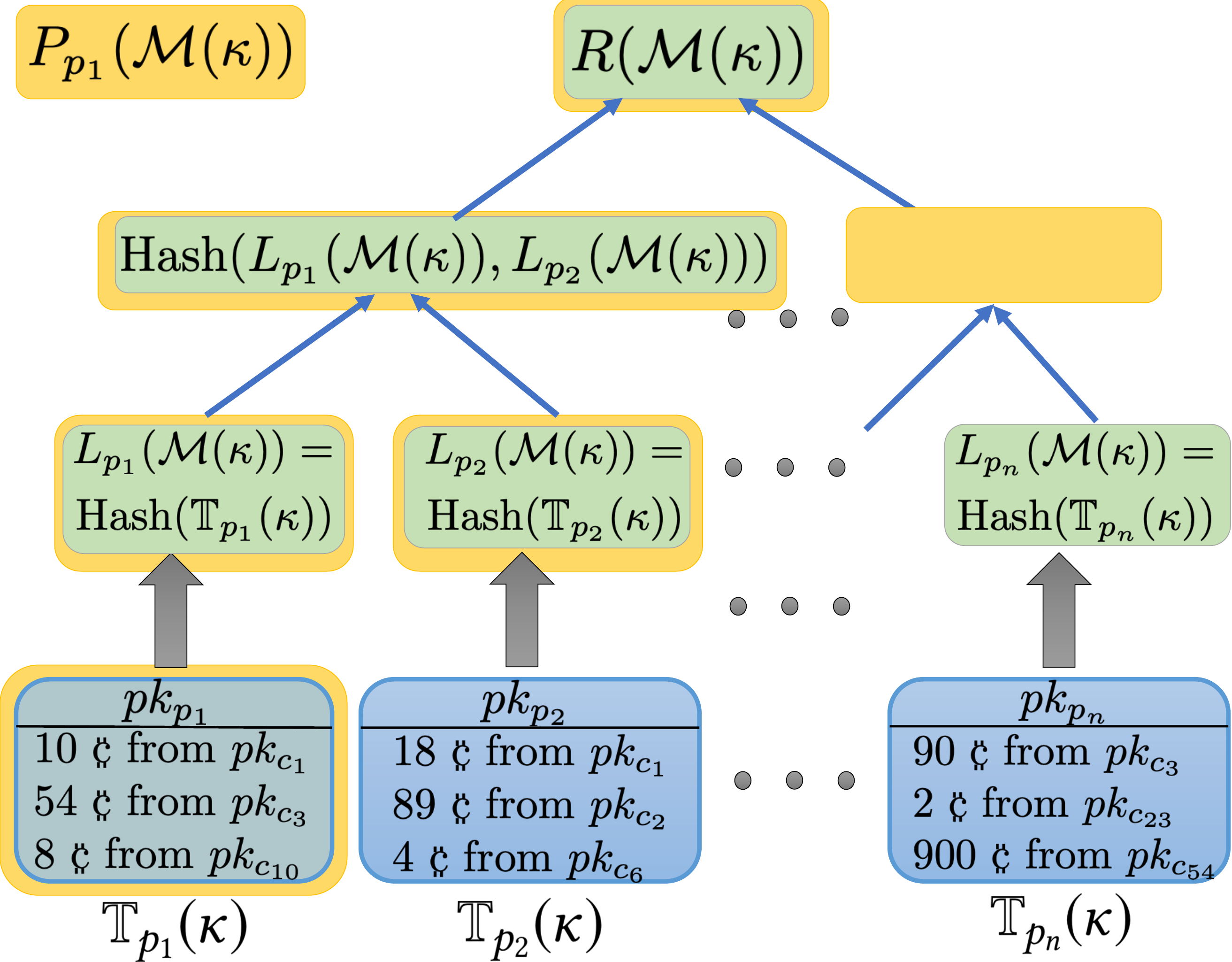}
    \caption{%
    Transactions in $\mathbb{T}_p$ reflect payments that $p$ is owed from different consumers. $\mathbb{T}_p$ is hashed to a leaf node $L_{p}(\mathcal{M})$ in the Merkle tree. A succinct Merkle proof of inclusion $P_{p}(\mathcal{M}(\kappa))$ for $\mathbb{T}_p(\kappa)$ can be given using the values of nodes with yellow borders.}
    \label{fig:plasmablock}
\end{figure}

\section{Numerical Example of the \name{} Model}
\label{sec:numericalExample}
To illustrate the consumer and provider transaction models in \name{}, consider $t = t_1, \dots, t_8$. Consumers $c_1$ and $c_2$ deposit $30$\currency{} and $20$\currency{} respectively into the \name{} smart-contract at time $t_1$. Suppose $c_1$ additionally deposits another $40$\currency{} at $t_3$. 
Then, $D_{c_1,t} = 30$ $\forall t \in [t_1, t_3)$ and $= 70$ $\forall t \in [t_3, t_8]$, and $D_{c_2,t} = 20$ $\forall t \in [t_1, t_8]$. To send $10$\currency{} to the operator at $t_4$, $c_1$ generates a transaction $T_a = (10, pk_\omega, pk_{c_1})$ and signature $\sigma_{T_a} = \mathcal{S}(T_a, sk_{c_1})$; then $\mu_{c_1,t_4}^* = 10$.  
Suppose $c_2$ similarly sends $10$\currency{} to the operator at $t_4$ via transaction $T_b$. To send another $10$\currency{}
at $t_5$, $c_1$ generates a transaction $T_c = (20, pk_\omega, pk_{c_1})$ and the corresponding $\sigma_{T_c}$; then $\mu_{c_1,t_5}^* = 20$. At $t_5$ then, the latest transactions $\mathbb{C}_{t_5} = \{T_b, T_c\}$ and the remaining funds available for consumers $c_1$ and $c_2$ to use at $t_5$ is $f_{c_1, t_5} = 50$ and $f_{c_2, t_5} = 10$ respectively. %
Suppose two registered providers $p_1$ and $p_2$ participate in the system, with no operator fees, and the orders for $T_a$ and $T_b$ are fulfilled by $p_1$ while $T_c$ is fulfilled by $p_2$. 
At $t_8$, the operator generates the block $\kappa = (\mathbb{T}, \mathcal{M})$, where $\mathbb{T}(\kappa) = \{ \{(10, pk_{p_1}, pk_{c_1}), (10, pk_{p_1}, pk_{c_2})\}, \{(10, pk_{p_2}, pk_{c_1})\}\}$
(with $\mathbb{T}_{\omega}(\kappa) = \emptyset$ since the operator did not deduct any fees in this case). 
We then have $\mathbb{T}_{p_1}(\kappa) = \{ \{(10, pk_{p_1}, pk_{c_1}), (10, pk_{p_1}, pk_{c_2})\}$, $\mathbb{T}_{p_2}(\kappa) = \{(10, pk_{p_2}, pk_{c_1})\}$, and $\mathbb{T}_{\omega}(\kappa) =  \emptyset$. Further, the Merkle tree $\mathcal{M}(\kappa) = H(H(\mathbb{T}_{p_1}(\kappa)), H(\mathbb{T}_{p_2}(\kappa)), H(\mathbb{T}_{\omega}(\kappa)))$, where $H$ is the one-way irreversible hash function used for generating the Merkle tree and the leaves of $\mathcal{M}$ are $L_p(\mathcal{M})=H(\mathbb{T}_p(\kappa))$, $p = p_1, p_2, \omega$. Suppose $pk_{p_1}$ withdraws their specified funds at $t_{10}$. Further, suppose the operator generates another block at $t_{16}$ and neither $pk_{p_1}$ nor $pk_{p_2}$ fulfil any additional consumer orders from $t_8$ to $t_{16}$. Then the generated block at $t_{16}$ is $\kappa' = (\mathbb{T}', \mathcal{M}')$, where $\mathbb{T}(\kappa') = \{(10, pk_{p_2}, pk_{c_1})\}\}$
(with $\mathbb{T}_{\omega}(\kappa)$ and $\mathbb{T}_{p_1}(\kappa)$ as $\emptyset$).

\section{Numerical Example of Usage of $\mathbb{M}$ and $\mathbb{N}$}
\label{app:MNExample}
We illustrate an example usage of states $\mathbb{M}$ and $\mathbb{N}$. Suppose $\mathbb{M}(-1) = \{pk_{a}\}$, $\mathbb{M}(-2) = \{pk_{b}, pk_{c}, pk_{d}\}$ and that $pk_{a}$, $pk_{b}$, $pk_{c}$, $pk_{d}$ had $10$\currency{}, $30$\currency{}, $40$\currency{}, $50$\currency{} sourced from $pk_{c1}$,$pk_{c5}$,$pk_{c1}$,$pk_{c5}$ respectively. Then $\mathbb{N}(-1) = \{(pk_{c1}, 10)\}$, $\mathbb{N}(-2) = \{(pk_{c1}, 40), (pk_{c5}, 80)\}$. Further, $\mathbb{N}(-1)^{pk_{c1}} = 10$\currency{}, $\mathbb{N}(-2)^{pk_{c1}}=40$\currency{} and $\mathbb{N}(-2)^{pk_{c5}} = 80$\currency{}. Finally, $\mathbb{M} = \{pk_{a}, pk_{b}, pk_{c}, pk_{d}\}$ in this example. 

\section{Rogue Key Attack in BLS Signature Aggregation}
\label{sec:rogueKey}
 
Consider a set of $n$ keys $K = \{(sk_i, pk_i) : 1 \leq i \leq n\}$ whose public keys and signatures are to be aggregated via BLS. An attacker who knows the public keys in $K$ can choose some $\beta \in \mathbb{Z}_q$ where $q$ is of prime order and compute a false public key $pk_{\text{att}} = g_1^\beta*  (\prod_{u=1}^{n}pk_u)^{-1}$ where $g_1$ is a generator for group $\mathbb{G}_1$ of prime order $q$. %
The aggregate public key computed by a verifier is $pk_a=pk_{\text{att}}*\prod_{u=1}^{n}pk_u$. The attacker can then declare the signature $\sigma_{a, m}=H_0(m)^\beta$ (where $H_0$ is a random oracle mapping into $\mathbb{G}_0$ which is also of prime order $q$) and convince the verifier that this has been signed by all $n$ $sk_i$'s as well as $pk_{\text{att}}$'s $sk_{\text{att}}$. To see this, note that verification of $\sigma_{a, m}$ requires checking if $e(g_1, \sigma_{a, m}) = e(pk_{\text{att}}*\prod_{u=1}^{n}pk_u, H_0(m))$ where $e$ is a pre-specified non-degenerate bilinear function ($e: \mathbb{G}_0 \times \mathbb{G}_1 \xrightarrow{} \mathbb{G}_T$). But $e(g_1, \sigma_{a, m}) = e(g_1, H_0(m)^\beta)$ as declared by the attacker to the verifier, and $ e(g_1, H_0(m)^\beta)=e(g_1^\beta, H_0(m))$, and $g_1^\beta = pk_{\text{att}}*\prod_{u=1}^{n}pk_u$ by definition of $pk_{\text{att}}$.

\section{Optimizing for Storage in \name{}}
\label{sec:storageOptimization}
\name{} can be easily modified to move the bulk of its state (consumer information) off-chain. %
The \name{} smart-contract stores the public key $pk_c$ of each consumer $c$ along with their net deposited funds $D_{c,t}$ on the contract at any time $t$, the highest operator-owned balance in their channel revealed to the contract as of time $t$, $\mu_{c,t}'$; and the total funds $w_{c,t}^*$ withdrawn by providers and the operator from this channel. Clearly, the required storage is $O(n)$ in the number of registered consumers (and $O(1)$ in the number of registered providers). If storage is significantly more expensive than compute on the underlying DLT, we can optimize for storage at higher computational cost. Similar to the storage mechanism proposed in ZK Rollup~\cite{zkrollup}, the contract can track just two Merkle roots instead of individual consumer data: one for a Merkle tree $A$ of registered consumers' public keys $pk_c$; and the other for a Merkle tree $B$ of tuples $(D_{c,t}, \mu_{c,t}', w_{c,t}^*)$, such that $pk_c$ that is stored in the $n$-th leaf of $A$ corresponds to $(D_{c,t}, \mu_{c,t}', w_{c,t}^*)$ in the $n$-th leaf of $B$. Consumer registrations can use the same process as ZK Rollup~\cite{zkrollup}; however, note that these functions are now more compute-intensive since each requires the contract to verify Merkle proofs. Similarly, provider and operator withdrawals require computation to verify the Merkle proofs for the consumer's balance in $B$ against which they attempt withdrawal. However, all changes to $B$ can be broadcast to others (stored in blockchain's logs), hence all providers and operators can compute the leaves and Merkle proofs in $B$ for any consumer, even if they have been offline.

\section{Lack of Support for BLS Account Keys on the Root-chain}
\label{app:nonBLSaccount}
Note that if the root-chain does not support BLS account keys (i.e., the keys used for signing blockchain transactions do not support BLS operations), the \vendor{} must explicitly specify their BLS public key $pk_p$ to the smart-contract during registration and provide a PoP for it. This PoP is denoted by $\sigma_{p, BLS} = \mathcal{S}(m$=$(pk_p,pk_p'), sk_{p})$, where $pk_{p}'$ is $p$'s root-chain account key. Since the account credentials on the root chain are different from the ones used for signing in \name{}, the \vendor{} must provide $\sigma_{p, BLS}$ for any transaction (e.g., withdrawals) with the smart-contract to prove its identity.

\section{Algorithm for Block Generation by the \name{} Operator}
\label{app:opAlg}
\begin{algorithm}
    \SetKwInOut{Input}{Input}
    \SetKwInOut{Output}{Output}
    \DontPrintSemicolon
    \SetKwFunction{FgetTransaction}{getTransaction}

    \Input{Set $O_t$ %
    with elements of type $(\mu', pk_p,pk_c)$ reflecting the total value $\mu'$ of $c$'s orders from the last $\gamma$ timesteps fulfiled by $p$}%
    \Output{$\kappa$}
    \KwData{current time $t$, $\mathbb{K}_t(-1)$, $\mathbb{W}$, $\mathbb{B}$, $\mathbb{R}$}
    \For{$p \in \mathbb{R}$}
    {
        \eIf{$p \notin \mathbb{W} \cup \mathbb{B}$}
        {
            $\mathbb{T}_p = \mathbb{T}_p(\mathbb{K}_t(-1))$
        }
        {
            $\mathbb{T}_p=\{\}$
        }
        \For{$(T'=(\mu',pk_p, pk_c)) \in O_t$}
        {
            $T'' = $getTransaction$(s(T'), \mathbb{T}_p)$
                        \Comment{Pass by ref} \;
            \eIf{$T'' = \bot$}
            {
                $\mathbb{T}_p.insert(T')$
            }
            {
                $\mu(T'') = \mu(T'')  + \mu(T')$
            }
        }
    }
    $\mathbb{T} = \bigcup\limits_{p \in \mathbb{R}} \{\mathbb{T}_{p}\}$\;
    $\mathcal{M} =$ merklize$(\mathbb{T})$ \Comment{Generates a Merkle tree whose leaves are hashes of elements in the input set.}\; %
    $\kappa = (\mathbb{T}, \mathcal{M})$\;
    \KwRet $\kappa$\;%
    \SetKwProg{Fn}{Def}{:}{}
    \Fn{\FgetTransaction{$pk_c$, $\mathbb{T}_p$}}{
        \For{$T' \in \mathbb{T}_p$}
        {
            \If{$s(T') = pk_c$}
            {
                \KwRet $T'$ \Comment{Pass by ref}
            }
        }
        \KwRet $\bot$\;
    }
    \caption{Block Generation by the Operator}
    \label{alg:opComputingTree}
\end{algorithm}
As shown in the \textbf{Data} field of Algorithm~\ref{alg:opComputingTree}, the operator keeps track of $\mathbb{K}_t(-1)$, $\mathbb{W}$,$\mathbb{B}$ and the set $\mathbb{R}$ of registered \vendors{} in the system. 

\section{Proofs}
\label{sec:proofs}
We provide proofs associated with the results in Section~\ref{sec:modules}.
\subsection{Proof of Lemma~\ref{lem:useOwnKey}}
\label{sec:appProof}
\begin{sproof}
First, we show that a public key $pk_p$ can  be registered only by its owner $p$ (who knows $sk_p$). 
Let $p$'s public and private keys on the root chain (i.e. account keys that are used for signing blockchain transactions) be denoted by $pk_p'$ and $sk_p'$ respectively. Consider $(pk_p', sk_p') = (pk_p, sk_p)$, i.e. the root chain supports BLS account keys which can also be used with \name{}. By design of the underlying blockchain, miners process transactions only if the transaction is signed by the stated sender, in this case, $pk_p$; hence registering a provider implicitly provides Proof of Possession (PoP) to the blockchain.
However, this can be attacked with an oracle~\cite{ristenpart2007power} 
OMSign$({pk_i}, msg)$ that returns $msg$ signed by $sk_i$. In that case, the attacker with a maliciously computed $pk_{\text{att}}=g_1^\beta*(\prod_{i=1}^{n} pk_i)$ may provide a signed transaction on the root chain  by computing $\sigma_{m, \text{att}} = H_0(m)^{\beta}/\prod_{i=1}^{n}$OMSign$(pk_i,m)$ where $m$ is the transaction to be submitted to the blockchain calling the contract's registration function and registering $pk_{\text{att}}$ as a vendor, $i$ iterates over the registered providers, and some $\beta \in \mathbb{Z}_q$. In \name{}, however, this requires that all registered providers (even honest ones) sign this transaction asking for $pk_{\text{att}}$'s enrolment, %
which they have no reason or incentive to do. 

Consider the other case where $(pk_p', sk_p') \neq (pk_p, sk_p)$. By design of the underlying blockchain, miners process transactions only if the transaction is signed by the stated sender, in this case, $pk_p'$; hence registering a provider implicitly provides Proof of Possession (PoP) to the blockchain. It then suffices to show that $pk_p'$'s owner also owns $pk_p$. First, we note that the provided PoP $\sigma_{p,BLS}$ establishes that $pk_p$ is not a rogue public key. Even if OMSign is available, employing separate hash functions for signing POP messages and other messages~\cite{ristenpart2007power} guarantees resilience of the provided PoP to the rogue public-key attack. Second, since a provider $p$ generates this PoP by signing the combined hash of their root account public key $pk_p'$ and BLS key $pk_p$, if $\mathcal{V}(pk_p, H(pk_p, pk_p'), \sigma_{p, BLS}) = 1$, then $pk_p$'s owner is the owner of $pk_p'$.

Finally, note that the contract rejects the registration unless $\sigma_{\omega,p}$ was generated after the latest freeze period (using the provided timestamp $\tau_r$. Hence, $\sigma_{\omega,p}$ provided by the operator to $p$ for registration is only valid until the next freeze window begins. If $p$ registers successfully with  $\sigma_{\omega,p}$, it cannot register again (with $\sigma_{\omega,p}$) even in the current open window since $pk_p$ is added to $\mathbb{B}$. The only way $p$ can register multiple times is if a colluding operator, knowing that $p$ has already registered, waits for the next notarization to succeed (which clears $\mathbb{B}$) and generates $\sigma_{\omega,p}$ with the latest $\tau_r$ and $p$ again calls the registration function with this.
\end{sproof}
\subsection{Proof of Theorem~\ref{thm:trackingMissingProviders}}
\begin{sproof}
For ease of explanation, we set $\mathbb{X}_t = \emptyset$ in the provided commitment (the proof trivially extends to the case where  $|\mathbb{X}_t| > 0$). The contract necessitates $apk_{\text{active}}\cdot{}\prod_{p \in \mathbb{M}_t} = apk$ to proceed with the commitment verification. First note that  registration of a rogue public key is not possible here, as shown in Lemma~\ref{lem:useOwnKey}; hence the typical rogue public key attack on BLS signature aggregation is infeasible here.

If the operator omits a registered provider from $apk_\text{active}$ or $\mathbb{M}_t$, then $apk_{\text{active}}\cdot{}\prod_{p \in \mathbb{M}_t} \neq apk$. To omit a registered provider while also ensuring $apk_{\text{active}}\cdot{}\prod_{p \in \mathbb{M}_t} = apk$, the operator generates an unregistered key-pair  $(pk_{\text{att}}, sk_{\text{att}})$ and includes $\sigma_{\text{att}, \kappa}$ when generating $ars_{\kappa}$ (or may simply set $ars_{\kappa} = \sigma_{\text{att}, \kappa}$). Let $\mathbb{A}'_t \subseteq \mathbb{A}_t$ be the subset of active signing providers whose signatures the operator includes in $ars_{\kappa}$ along with $\sigma_{\text{att}, \kappa}$. The corresponding aggregate public key that will then verify $ars_{\kappa}$ successfully is $pk_{\text{att}}\cdot{}\prod_{p \in \mathbb{A}'_t} pk_p$. Hence, the operator provides $apk_{\text{active}} = pk_{\text{att}}\cdot{}\prod_{p \in \mathbb{A}'_t} pk_p$ to the contract. To satisfy the contract's requirement, the operator generates $pk_{\text{missing}} = apk\cdot{}(apk_{\text{active}})^{-1}$ and sets $\mathbb{M}_t = \{pk_{\text{missing}}\}$. This ensures that contract's check $apk_{\text{active}}\cdot{}\prod_{p \in \mathbb{M}_t} = apk$ passes; however, note that the contract also requires proofs of possession $\sigma_{p, \text{init}}$ from each $p \in \mathbb{M}_t$. To generate $\sigma_{\text{missing}, \text{init}}$ requires computing the secret key  $sk_{\text{missing}}$ given the public key $pk_{\text{missing}}$ (which requires violating the Diffie-Hellman assumption), while it can be trivially provided for legitimately missing providers who relayed their  $\sigma_{p, \text{init}}$ at the beginning to acquire $\sigma_{\omega, p}$ for registration. Even if the operator includes  legitimately missing providers in $\mathbb{M}_t$, the presence of a POP-less $pk_{\text{missing}}$ in $\mathbb{M}_t$ is imminent, which renders the attack unsuccessful. 
\end{sproof}

\subsection{Proof of Theorem~\ref{thm:excludingNonRegisteredProviders}}
\begin{sproof}
For ease of explanation, we set $\mathbb{X}_t = \emptyset$ in the provided commitment (the proof trivially extends to the case where  $|\mathbb{X}_t| > 0$). 
We show that inserting credentials that have not been registered with the \name{} smart-contract in $apk_{\text{active}}$ or $\mathbb{M}_t$ will cause the commitment to be rejected. 
Consider that the operator generates an unregistered key-pair  $(pk_{\text{att}}, sk_{\text{att}})$ and includes $\sigma_{\text{att}, \kappa}$ in generating $ars_{\kappa}$ (or may simply set $ars_{\kappa} = \sigma_{\text{att}, \kappa}$).
Let $\mathbb{A}'_t \subseteq \mathbb{A}_t$ be the subset of active signing providers whose signature the operator includes in $ars_{\kappa}$ along with $\sigma_{\text{att}, \kappa}$. 
The corresponding aggregate public key that will then verify $ars_{\kappa}$ successfully is $pk_{\text{att}}\cdot{}\prod_{p \in \mathbb{A}'_t} pk_p$. 
Hence, the operators provides $apk_{\text{active}} = pk_{\text{att}}\cdot{}\prod_{p \in \mathbb{A}'_t} pk_p$ to the contract. To satisfy the contract's requirement that $apk_{\text{active}}\cdot{}\prod_{p \in \mathbb{M}_t} = apk$, the operator generates $pk_{\text{missing}} = apk * (apk_{\text{active}})^{-1}$ and sets $\mathbb{M}_t = \{pk_{\text{missing}}\}$. 
This ensures that contract's check $apk_{\text{active}}\cdot{}\prod_{p \in \mathbb{M}_t} = apk$ passes; however, note that the contract also requires proofs of possession $\sigma_{p, \text{init}}$ from each $p \in \mathbb{M}_t$. 
To generate $\sigma_{\text{missing}, \text{init}}$ requires computing the secret key  $sk_{\text{missing}}$ given the public key $pk_{\text{missing}}$ (which requires violating the Diffie-Hellman assumption), while it can be trivially provided for legitimately missing providers who relayed their  $\sigma_{p, \text{init}}$ at the beginning to acquire $\sigma_{\omega, p}$ for registration. 
Even if the operator includes  legitimately missing providers in $\mathbb{M}_t$, the presence of a POP-less $pk_{\text{missing}}$ in $\mathbb{M}_t$ is imminent, which renders the attack unsuccessful. Finally, consider that the operator generates $ars_{\kappa} = \prod_{p \in \mathbb{A}_t}\sigma_{p, \kappa}$ and the corresponding $apk_{\text{active}} = \prod_{p \in \mathbb{A}_t} p$ correctly. It is straightforward to see that inserting unregistered key $pk_{\text{att}}$ in a correctly generated $\mathbb{M}_t$ will cause the commitment to be rejected since then $apk_{\text{active}}\cdot{}\prod_{p \in \mathbb{M}_t} \neq apk$. %

For credentials in $\mathbb{X}_t$ of an accepted notarization, note that the contract requires $\mathbb{X}_t \subseteq \mathbb{X}$, where $\mathbb{X}$ consists of \vendors{} who have exited or deregistered the system (using the contract's withdrawal module) since the last notarization. Hence $\mathbb{X}_t$ is guaranteed to consist of \vendors{} who had deregistered after the last notarization. If $\mathbb{X}_t$ consisted of \vendors{} who had not registered at an earlier time with the \name{} smart-contract, then $apk_{\text{active}}\cdot{}\prod_{p \in \mathbb{M}_t} \prod_{p \in \mathbb{X}_t} \neq apk$ based on the same reasoning above, in which case the contract would reject the commitment.
\end{sproof}
\subsection{Proof of Theorem~\ref{thm:consumerFundsSafety}}
\begin{sproof}
A consumer's confirmed funds $f_{c,t}$ at time $t$ is $D_{c,t} - \mu_{c,t}^\ast$; hence the cumulative funds withdrawn by \vendors{} or the operator against $c$'s deposit should not exceed $\mu_{c,t}^\ast$ at any time $t$. In other words, if $w_{c,t}^\ast$ denotes the total funds withdrawn against $c$'s deposit as of time $t$, then we need to show $w_{c,t}^\ast\le\mu_{c,t}^\ast$. As described in the withdrawal module,  $w_{c,t}^\ast$ is tracked and updated by the contract every time a withdrawal against $c$'s funds is made successfully. From Algorithm~\ref{alg:withdrawal}, we see that the maximum funds transferred upon a successful withdrawal by provider $p$ is min$\{\mu_{c,t}'-\sum_{i=\eta}^{x-1} \mathbb{N}(-i)^{pk_c} - w_{c,t}^\ast, \mu(T')\}$. Let $w_p$ denote this value. Even if no missing \vendors{} exist (i.e. $\mathbb{N} = \emptyset$), a maximum of only $w_p = \mu_{c,t}' - w_{c,t}^\ast$ is transferred to the \vendor{}; $w_{c,t+1}^\ast$ is then updated to $w_{c,t}^\ast + w_p = \mu_{c,t}'$ and $\mu_{c,t}' \leq \mu_{c,t}^\ast$ by definition. Since $w_{c,t+1}^\ast =  \mu_{c,t}'$, subsequent withdrawals will not transfer any funds out of $c$.

\end{sproof}
\subsection{Proof of Theorem~\ref{thm:providerSafety}}
\begin{sproof}
To show \textit{\vendorCap{} Safety}, we show that a \vendor{} following the \name{} protocol when signing a commitment is guaranteed safety of funds assigned therein to her even if she becomes arbitrarily unavailable to sign future commitments (or does not wish to, due to detecting malfeasance in subsequent operator-generated commitments). Since the operator is subject to the same rules as a \vendor{} for the registration and withdrawal process (and implicitly attests a commitment by generating it and submitting it to the smart-contract), showing Safety for \vendors{} also secures the Operator's funds in the commitment. We prove this by showing that the following two statements hold:

\begin{itemize}
    \item Case 1: Suppose $p$ signs a block $\kappa$'s commitment (i.e. the tuple $(R(\mathcal{M}(\kappa)), \tau)$) and $\kappa$'s notarization is published to the contract with $\sigma_{p,\kappa}$ included in $ars_{\kappa}$. Then $p$'s funds are safe as long as $\mathbb{K}_t(-1)=\kappa$, i.e. $p$'s funds in $\kappa$ are fully available for $p$ to withdraw as long as $\kappa$ is the latest commitment.
    \item Case 2: Suppose $p$'s signature is not included in the commitment of a block $\kappa'$ and $\kappa'$ is published to the contract (i.e. $\mathbb{K}_t(-1) = \kappa'$ at some time $t$). Define $x$ such that $\mathbb{K}_t(-x) = \kappa_p$, where $\kappa_p$ is the last notarized block that $p$ signed. Then funds assigned to $p$ in $\kappa_p$ ($\mathbb{T}_p(\kappa_p)$) are available for withdrawal by $p$ at $t$.
\end{itemize}

\textbf{Case 1.} \textbf{Part a)} We first show that $p$'s funds in a notarized block $\kappa$ that $p$ signed for at $t'$ is safe for any $t > t'$ as long as providers and the operator can only withdraw funds that have been assigned in $\kappa$.

Funds assigned to be $\mathbb{T}_p$ are of the form shown in Figure~\ref{fig:plasmablock}; for each consumer whose order $p$ fulfiled, it specifies a payment amount to be sourced from that consumer's deposit.
WLOG, assume that $p$ has funds assigned from exactly one consumer $c$ in $\mathbb{T}_p$; i.e. $|\mathbb{T}_p = 1|$ and $\mathbb{T_p} = {T'}$ where $s(T') = pk_c$. Let $T$ be the corresponding source transaction that is revealed to $p$ by the operator as part of $\mathbb{C}_{t'}$ for block verification and signing. Recall that the set of providers that each have a leaf in $\mathcal{M}(\kappa)$ is denoted by $\mathbb{P}(\kappa)$. Let $\mathbb{P}(\kappa)_{-p} = \mathbb{P}(\kappa) \setminus p$. Then it suffices to show that: (1) at time $t$ the contract has at least $\mu(T')$ amount available as $c$'s deposited funds $D_{c,t}$, and (2) that the total funds that can be withdrawn by $\mathbb{P}(\kappa)_{-p}$ cannot exceed $D_{c,t} - \mu(T')$. 
To prove (1) Note that by definition of \name{} (Algorithm~\ref{alg:providerSigningCommitment}), $\mu(T) \leq D_{c,t'}$ and $\sum_{v \in \mathbb{P}(\kappa)} \sum_{T'' \in \mathbb{T}_v(\kappa)} \mathbbm{1}_{s(T'') = c} \,\, \mu(T'') \leq \mu(T) - w_{c,t}^*$, implying $\mu(T')\leq\mu(T)\leq D_{c,t}$. Note that $D_{c,t}$ is a weakly monotonically increasing function of $t$ since consumer withdrawals are prohibited in \name{}; hence (1) holds. To prove (2), first note that a \vendor{} can withdraw funds in $\kappa$ only once as they are then tracked in $\mathbb{X} \cup \mathbb{W}$ and further withdrawals against $\kappa$ cancelled (Line 5-6 of Algorithm~\ref{alg:withdrawal}) . Let $\mu_{-p}$ be the total funds assigned to $\mathbb{P}(\kappa)_{-p}$ with $c$ as the source consumer. By definition, when $p$ signed $\kappa$ at $t'$, the following held: $\mu_{-p} + \mu(T') \leq \mu(T) \leq D_{c,t'}$. Subtracting $\mu(T')$ from this, we get $\mu_{-p} \leq \mu (T) - \mu(T') \leq D_{c,t}-\mu(T')$. However, this is contradictory if $\mu_{-p} > D_{c,t'} - \mu(T')$. Since $D_{c,t}$ monotonically increases with $t$ as well (weakly), this proves (2). 

\textbf{Part b)} 
To show Case 1 then, it suffices to show that as long as $\mathbb{K}_t(-1) = \kappa$, a withdrawal initiated by any \vendor{} $p'$ corresponds to funds accounted for $p'$ in $\kappa$. By design of the withdrawal function (Algorithm~\ref{alg:withdrawal}), if $p' \notin \mathbb{M}$, then $p$ can only withdraw funds assigned to her in $\mathbb{K}_t(-1)$ (i.e. the contract requires a Merkle proof showing that the submitted transaction set $\mathbb{T}_{p'}$ is included in $\mathcal{M}(\mathbb{K}_t(-1)$. 
If $p' \in \mathbb{M}$, note that the withdrawal function only allows $p'$ to withdraw funds stated in $\mathbb{T}_{p'}(\kappa_{p'})$ (i.e. the last notarized block $p'$ signed). If $ p \in \mathbb{M}(-x)$ for $x \ge 2$ (i.e. $p'$ had missed signing $\mathbb{K}_t(-2)$ as well), then Lines 14-16 of Algorithm~\ref{alg:providerSigningCommitment} ensures that $p$ does not sign $\mathbb{K}_t(-1)$ unless $\mathbb{T}_{p'}(\kappa_{p'})$ is included in $\mathbb{K}_t(-1)$. If, instead if $p' \in \mathbb{M}(-1)$, then the commitment verification module that notarized $\mathbb{K}_t(-1)$ enforced that $\mathbb{T}_{p'}(\mathbb{K}_t(-1)) \supseteq \mathbb{T}_{p'}({\mathbb{K}_t(-2})$ (Lines 19-29 in Algorithm~\ref{alg:commitmentVerification}). Hence this reduces to the proof of Part a) above.
Note that the timing constraints in \name{} (i.e. the checks in Lines 1-2 of the commitment generation Algorithm~\ref{alg:commitmentVerification}) ensures that older blocks signed by $p$ cannot be re-notarized by the operator.

\textbf{Case 2:} From Theorem~\ref{thm:trackingMissingProviders}, $p$ is detected as missing in any commitment that does not contain $p$'s signature (since $p$ has registered). 
When the contract processes the commitment for $\mathbb{K}_t(-x+1)$ (i.e. the first commitment submitted without $p$'s signature after the last notarized block containing $p$'s signature), 
the funds in $\mathbb{T}_p(\mathbb{K}_t(-x))$ are recorded in $\mathbb{N}$ against each source consumer, based on the definition of $\mathbb{N}$. Hence, (1) since $p$ is detected as missing, $p$'s confirmed funds (i.e. funds in the last block $\kappa_p = \mathbb{K}_t(-x)$ signed by $p$) are tracked in $\mathbb{N}$.
Further, note that the contract updates $\mathbb{N}$ to remove release the reservation of these funds only when $p$ is detected as having signed a submitted commitment again (Line $1-6$ of Algorithm~\ref{alg:commitmentVerification} ensures that the contract is provided the original transaction set $\mathbb{T}_p(\mathbb{K}_t(-x))$ again in that case to perform the update).
Next, (2) no \vendor{} who has signed a commitment that was subsequently published after $\mathbb{K}_t(-x)$ can withdraw funds \textit{locked by $\mathbb{N}(-y)$ for $y \in [x,\eta]$} assigned to \vendors{} whose last signature was on $\mathbb{K}_t(-x)$ or earlier. The maximum allowed withdrawal for a \vendor{} $p'$ in Algorithm~\ref{alg:withdrawal} Line 19 ensures this: $\text{min}\{\text{max}\{\mu_{c,t}' - \sum_{i = \eta}^{\text{blocksP'Missed}-1} \mathbb{N}(-i)^{pk_c} - w_{c,t}^*, 0\}, \mu'\}$  for each $T' \in \mathbb{T}_{p'}$ with $c = s(T')$. 
Finally, (3) at the time $t'$ that $p$ signed $\mathbb{K}_t(-x)$, note that funds reserved so far in $\mathbb{N}$, i.e. $\sum_{i = \eta}^{x-1}\mathbb{N}(-i)^{pk_c}, \forall c$ have already been incorporated (accounted for) in $\mathbb{K}_t(-x)$. Lines $15-16$ of Algorithm~\ref{alg:providerSigningCommitment} executed by \vendors{} for commitment signing ensures this. Hence the double-spend verification checks performed by $p$ in Lines 17-20 of Algorithm~\ref{alg:providerSigningCommitment} ensures that funds assigned to $p$ are not double-spent against the funds already reserved in $\mathbb{B}$ (cf. Case 1 above). Note that confirmed funds of a non-signing \vendor{} is reserved in $\mathbb{N}$ only the first time that their signature is detected missing after being present on the last notarization; subsequent commitments with the \vendor{}'s signature continuing to be absent does not result in modifications to $\mathbb{N}$. 
\end{sproof}

\subsection{Explanation of Corollary~\ref{cor:dataAvailability}}
Given Theorem~\ref{thm:providerSafety}, \textit{Income Certainty} is straightforward to infer from Algorithm~\ref{alg:commitmentVerification} and Algorithm~\ref{alg:withdrawal}. \textit{Data Availability} follows by design of the Withdrawal Function. A \vendor{} $p$'s confirmed funds $f_{p,t}$ in \name{} corresponds to the funds assigned to them in the last notarized block they signed, $\mathbb{T}_p(\kappa_p)$. If $p$ signed the last notarized block, Algorithm~\ref{alg:withdrawal} expects a Merkle proof for $\mathbb{T}_p(\kappa_p)$ against $R(\mathcal{M}(\mathbb{K}_t(-1))$, which $p$ knows since $p$ signed $\mathbb{K}_t(-1)$. If $p$ did not sign the last notarized commitment, then the contract simply expects the transaction set they last signed for, $\mathbb{T}_p(\kappa_p)$. Since $p$ is detected as missing in the first commitment she misses after the last notarization that had her signature (Theorem~\ref{thm:trackingMissingProviders}), (Algorithm~\ref{alg:commitmentVerification}, Lines 25-26 ensures that the contract is provided $\mathbb{L}_p$ by the operator when $p$ is missing to store  $H(\mathbb{T}_p(\kappa_p))$ in $\mathbb{L}_p$), as long as $p$ provides  $\mathbb{T}_p(\kappa_p)$, the check in Line 8 of Algorithm~\ref{alg:withdrawal} succeeds.
\appendices